\pdfoutput=1
\documentclass{ar-1col}
\usepackage[numbers]{natbib}
\usepackage{bm}
\usepackage[table, dvipsnames]{xcolor}
\definecolor{shadecolor}{cmyk}{0.03,0.03,0.12,0.0}
\renewcommand{\vec}[1]{\bm{#1}}

\jname{Xxxx. Xxx. Xxx. Xxx.}
\jvol{AA}
\jyear{YYYY}
\doi{10.1146/((please add article doi))}

\begin{document}

\markboth{Huang and Hoffman}{Monolayer FeSe on SrTiO$_3$}

\title{Monolayer FeSe on SrTiO$_3$}

\author{Dennis Huang$^1$ and Jennifer E. Hoffman$^{1, 2}$
\affil{$^1$Department of Physics, Harvard University, Cambridge, Massachusetts 02138; emails: dhuang@physics.harvard.edu, jhoffman@physics.harvard.edu}
\affil{$^2$Department of Physics \& Astronomy, University of British Columbia, Vancouver, British Columbia V6T 1Z1, Canada}}

\begin{abstract}
Epitaxial engineering of solid-state heterointerfaces is a leading avenue to realizing enhanced or novel electronic states of matter. As a recent example, bulk FeSe is an unconventional superconductor with a modest transition temperature ($T_c$) of 9 K. When a single atomic layer of FeSe is grown on SrTiO$_3$, however, its $T_c$ can skyrocket by an order of magnitude to 65 K or 109 K. Since this discovery in 2012, efforts to reproduce, understand, and extend these findings continue to draw both excitement and scrutiny. In this review, we first present a critical survey of experimental measurements performed using a wide range of techniques. We then turn to the open question of microscopic mechanisms of superconductivity. We examine contrasting indications for both phononic (conventional) and magnetic/orbital (unconventional) means of electron pairing, and speculations about whether they could work cooperatively to boost $T_c$ in a monolayer of FeSe.
\end{abstract}

\maketitle

\tableofcontents

\section{INTRODUCTION}

Interface engineering lies at the vanguard of current research in condensed matter physics and novel materials. From a fundamental perspective, quantum-size and electron correlation effects are enhanced in reduced dimensionality, often resulting in unexpected phenomena. From a technological perspective, as the pace of electronics miniaturization fast approches the limit of conventional semiconductors, alternative paradigms, such as atomically-thin materials and interfaces with manifestly quantum behavior, are needed to assume the mantle of next-generation electronics. With improving ability to assemble atomically-sharp interfaces ``bottom up'' through molecular beam epitaxy (MBE) or mechanical stacking of 2D-layered materials, many possibilites abound.

As the central example of this review, interface engineering holds promise in tuning, boosting, or generating superconducting states of matter -- low-temperature quantum phases in which electrons form Cooper pairs and charge transport is lossless. Since its discovery in 1911, superconductivity has continued to fascinate and baffle condensed matter physicists, while the goal of realizing room-temperature superconductivity remains elusive. Within the past decade, various examples of interface superconductivity have been observed. When two insulating oxides, LaAlO$_3$ and SrTiO$_3$, are put togther, a superconducting electron gas is formed at the interface, albeit with a low transition temperature ($T_c$) of 200 mK~\cite{Reyren_Science_2007}. When a bilayer of insulating La$_2$CuO$_4$ and metallic La$_{1.55}$Sr$_{0.45}$CuO$_4$ is formed, the aggregate system displays a $T_c$ exceeding 50 K~\cite{Gozar_Nat_2008}. And as the latest example, when a single-unit-cell (1UC) layer of FeSe is deposited on SrTiO$_3$~\cite{Wang_CPL_2012}, its $T_c$ skyrockets up to 65 K~\cite{He_NatMat_2013, Tan_NatMat_2013, Lee_Nat_2014, Zhang_SciBull_2015} or 109 K~\cite{Ge_NatMat_2015}, an order of magnitude above its bulk $T_c$ of 9 K. In this review, we will cover key experimental and theoretical developments related to 1UC FeSe/SrTiO$_3$ up to early 2016. We focus on measurements of basic properties and questions of superconducting mechanisms.

\subsection{Approaching the 2D limit with FeSe}

\begin{figure}[h]
\includegraphics{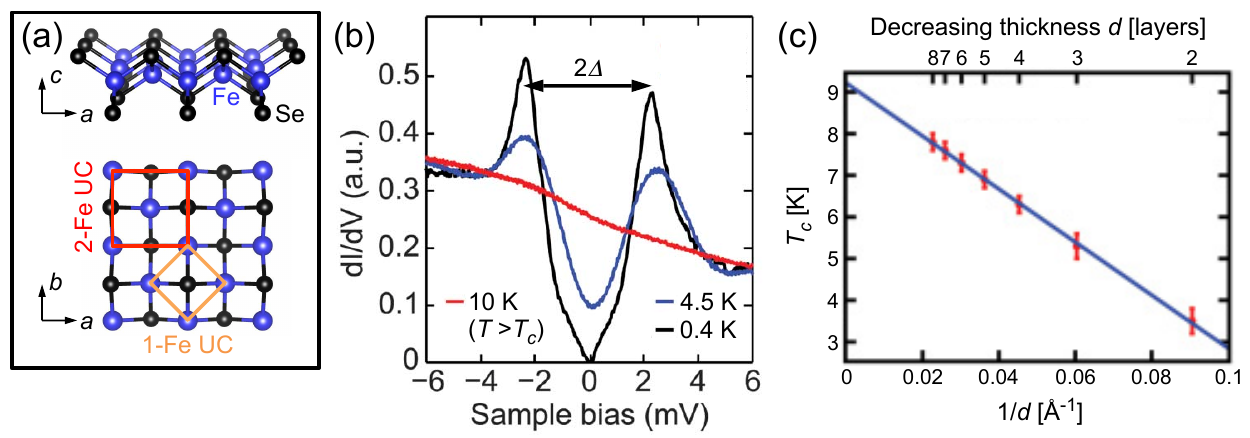}
\caption{(\textit{a}) Crystal structure of an FeSe monolayer; side and top views. The orange and red boxes enclose the 1-Fe UC and 2-Fe UC, respectively. (\textit{b}) STM $dI/dV$ spectra of multilayer FeSe/SiC, exhibiting a V-shaped superconducting gap of $\Delta$ = 2.2 meV at $T$ = 0.4 K, which disappears above 10 K. Adapted from Ref.~\cite{Song_Science_2011}. (\textit{c}) Gap-closing temperature, $T_c$, of multilayer FeSe/SiC as a function of inverse FeSe thickness, $1/d$. Adapted from Ref.~\cite{Song_PRB_2011}.}
\label{Fig2D}
\end{figure}

FeSe posesses the simplest structure among the iron-based superconductors, consisting of superconducting Se-Fe-Se triple layers stacked by van der Waals forces, with no buffer layers~\cite{Hsu_PNAS_2008}. \textbf{Figure~\ref{Fig2D}\textit{a}} shows the structure of one triple layer, which includes Fe atoms arranged in a square lattice and Se atoms staggered above and below the Fe-plane. Due to the staggering, the primitive UC contains two Fe atoms (and two Se atoms). However, since the low-energy bands of FeSe are dominated by Fe $3d$ orbitals, many theories or spectroscopies reference the 1-Fe UC for convenience. 

By virtue of its structural simplicity, FeSe should be the prototypical iron-based superconductor to investigate, except it proved difficult to synthesize in high quality at first. Its superconducting polymorph occupies a narrow region in the Fe-Se alloy phase diagram~\cite{Okamoto_JPE_1991}, complicating common melt and self-flux growths. In 2011, Song \textit{et al.} used MBE to grow pristine FeSe films on graphitized SiC~\cite{Song_Science_2011, Song_PRB_2011}. Using scanning tunneling microscopy (STM), they resolved clean surfaces with only one atomic defect per 70,000 Se sites. Measurements of tunneling conductance ($dI/dV$), which is proportional to the local density of states, revealed two signatures of a superconducting state: (1) A V-shaped gap of $\Delta$ = 2.2 meV, representing the binding energy of paired electrons, that disappeared above 10 K (\textbf{Figure~\ref{Fig2D}\textit{b}}); (2) vortices in the presence of a perpendicular magnetic field. Although MBE-grown films are not amenable to many bulk and thermodynamic probes, they have other advantages. Both the monolayer limit and interface interactions with different substrates can be readily examined.

Song \textit{et al.} found that the FeSe films interacted weakly with the graphitized SiC substrate (islands could be displaced by an STM tip), and were thereby close to the free-standing limit~\cite{Song_PRB_2011}. Upon decreasing film thickness, $T_c$, as measured by the gap-closing temperature, dropped from 7.8 K (8UC-thick FeSe) to below 2.2 K (1UC-thick FeSe), the base temperature of their experiment (\textbf{Figure~\ref{Fig2D}\textit{c}}). The drop exhibited a $1 - d_c/d$ dependence, $d$ being the film thickness and $d_c$ being a critical value. This thin-film behavior was explained long ago as resulting from a general, surface boundary condition with the Ginzburg-Landau equation~\cite{Simonin_PRB_1986}. Thus in 2011, it did not appear that the 2D limit of FeSe would be a promising regime to explore, unless new microscopic effects could be introduced.

\subsection{\label{boost}Monolayer FeSe gets an oxide boost}

\begin{figure}[h]
\includegraphics{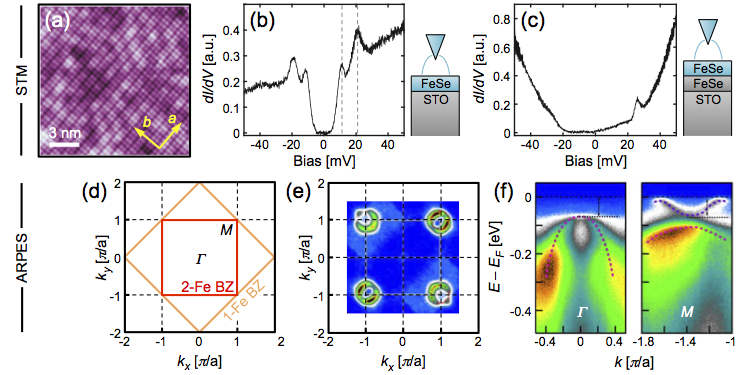}
\caption{(\textit{a})-(\textit{c}) Initial STM measurements of 1UC FeSe/SrTiO$_3$, adapted from Ref.~\cite{Wang_CPL_2012}. (\textit{a}) Atomically-resolved topography. Each bright spot represents a top-layer Se atom. (\textit{b}), (\textit{c}) Contrasting $dI/dV$ spectra of 1UC (superconducting) and 2UC (non-superconducting) FeSe/SrTiO$_3$. The dashed vertical lines in (\textit{b}) mark two gap edge peaks at 9 mV and 20.1 mV. The schematics illustrate that the tunneling depth is largely restricted to the surface FeSe layer, so it is not possible to determine from (\textit{c}) alone whether the presence of the second UC has altered the high-$T_c$ superconductivity in the first UC. (\textit{d})-(\textit{f}) Initial ARPES measurements of 1UC FeSe/SrTiO$_3$, adapted from Ref.~\cite{Liu_NatComm_2012}. (\textit{d}) Brillouin zone (BZ) conventions. (\textit{e}) FS map, revealing electron pockets at the BZ corner $M$ and the overall electron-doped nature of 1UC FeSe/SrTiO$_3$. (\textit{f}) High-symmetry cuts across the BZ center $\Gamma$ and corner $M$, revealing additional occupied bands.}
\label{FigFY}
\end{figure}

It came as a great surprise a year later that monolayer FeSe could undergo an order-of-magnitude $T_c$ enhancement when grown epitaxially on SrTiO$_3$(001). The lattice mismatch between bulk FeSe (\textit{a} = 3.77 \AA~\cite{Bohmer_PRB_2013}) and SrTiO$_3$ (\textit{a} = 3.905 \AA~\cite{Schmidbauer2012_ACSB_2012}) is roughly 3\%. STM measurements by Wang \textit{et al.} revealed a topographic period-doubling (\textbf{Figure~\ref{FigFY}\textit{a}}) and a large U-shaped, double-gap structure (9.0 meV and 20.1 meV) in 1UC FeSe/SrTiO$_3$ (\textbf{Figure~\ref{FigFY}\textit{b}}), with closing temperature $T_c$ above their experimental limit of 42.9 K. Intriguingly, this superconductivity boost did not persist or even proximitize low-$T_c$ superconductivity in a second UC of FeSe deposited on the heterostructure. STM $dI/dV$ measurements, whose probing depth is likely limited to the surface FeSe layer, instead showed a semiconducting spectrum on the second FeSe layer (\textbf{Figure~\ref{FigFY}\textit{c}}). This observation points to an underlying interface effect, one that is atomically localized to the first UC of FeSe on SrTiO$_3$. Wang \textit{et al.} speculated that electron-phonon coupling could be enhanced at the interface and boost $T_c$, based on their previous work with Pb/Si(111) and In/Si(111) films~\cite{Zhang_NatPhys_2010}.

Due to technical challenges, Wang \textit{et al.} could measure transport only in a Si-capped, 5UC FeSe/SrTiO$_3$ heterostructure. They measured zero resistance at some temperature lower than 30 K, and extrapolated a resistive onset temperature around 53 K. (As shown by STM spectroscopy in \textbf{Figure~\ref{FigFY}\textit{b,c}}, the superconducting signal originates from the interface FeSe layer only.) 

Angle-resolved photoemission spectroscopy (ARPES) measurements in the same year provided initial insights into the role of the interface. ARPES can map filled-state band structure in momentum space. Liu \textit{et al.}~\cite{Liu_NatComm_2012} found that the Fermi surface (FS) of 1UC FeSe/SrTiO$_3$ comprises nearly-circular electron pockets at the Brillouin zone (BZ) corner $M$ (\textbf{Figure~\ref{FigFY}\textit{d,e}}). In contrast to bulk FeSe~\cite{Shimojima_PRB_2014, Nakayama_PRL_2014, Watson_PRB_2015, Zhang_PRB_2015}, where additional hole FSs exist at the zone center $\Gamma$, here the corresponding hole pocket is sunken 65-80 meV below the Fermi energy ($E_F$) (\textbf{Figure~\ref{FigFY}\textit{f}}). Assuming doubly-degenerate electron pockets, a Luttinger count yields 0.10 electrons/Fe atom. Thus, relative to its bulk, 1UC FeSe appears to be electron-doped from the substrate. To provide further support for the superconducting nature of 1UC FeSe/SrTiO$_3$, Liu \textit{et al.} resolved nearly-isotropic gaps on the electron pockets at each $M$ point, of values 13$\pm$2 meV and 15$\pm$2 meV for two samples. They found the gap-closing temperature to be 55$\pm$5 K.

Before proceeding, we reiterate that monolayer FeSe/SrTiO$_3$ is not monolayer FeSe. A giant $T_c$ enhancement is present only in the former, due to some effect introduced by the SrTiO$_3$.

\section{EXPERIMENTAL CHALLENGES}

A foremost challenge related to 1UC FeSe/SrTiO$_3$ has been the characterization of its growth, atomic structure, and superconducting metrics. As a point of emphasis, bulk probes are not effective for this system. Not only is the cross section of a single UC layer miniscule, but also FeSe exhibits extreme air sensitivity, hampering \textit{ex-situ} measurements. Thus, the basic goal of determining $T_c$ represents a nontrivial endeavor requiring customized and integrated instrumentation in ultra-high-vacuum. Example apparatuses include combined MBE-ARPES-STM systems, double chalcogen-MBE/oxide-MBE chambers, and customized \textit{in-situ}, four-point probes. 

In this section, we review various experiments related to film characterization, categorized under three questions: What is $T_c$? What are the necessary growth conditions? What is the interface structure? We attempt to reflect the sentiments of the scientific community by conveying both the excitement related to the spectacular findings of \textit{tour-de-force} experiments, and scrutiny related to the challenging nature of these feats and of film quality/homogeneity.

\subsection{What is $T_c$?}

\textbf{Table~\ref{TabTc}} presents a comparison of $T_c$ measurements across different probes, heterostructures, and laboratories. Among various \textit{in-situ} ARPES measurements~\cite{He_NatMat_2013, Tan_NatMat_2013, Lee_Nat_2014}, there is consensus in a gap-closing temperature $T_c$ $\sim$65 K. Some variation exists with the degree of post-growth annealing~\cite{He_NatMat_2013} (see Subsection~\ref{growth} for details). Enhancement of $T_c$ up to 75 K is possible if extra tensile strain is introduced through an additional KTaO$_3$ substrate~\cite{Peng_NatComm_2014}. 

\begin{table}[h]
\footnotesize
\caption{Comparison of $T_c$ measurements across different probes, heterostructures, and laboratories. We distinguish measurements without (\textit{in situ}) and with (\textit{ex situ}) a capping layer.}
\label{TabTc}
\begin{center}
\begin{tabular}{c|c|c|c|c}
\hline
\cellcolor{shadecolor}\textbf{Technique} & \cellcolor{shadecolor}\textbf{Definition} & \cellcolor{shadecolor}\textbf{Heterostructure} & \cellcolor{shadecolor}\textbf{Value [K]} & \cellcolor{shadecolor}\textbf{Ref.}  \\ 
\hline
\textbf{\textit{in-situ}} & & & & \\
STM & Gap-closing $T$ & 1UC FeSe/Nb:SrTiO$_3$ & $>$42.9 & \cite{Wang_CPL_2012} \\
STM & Gap-closing $T$ & 1UC FeSe/SrTiO$_3$ & $>$50.1 & \cite{Zhang_PRB_2014} \\
ARPES & Gap-closing $T$ & 1UC FeSe/Nb:SrTiO$_3$ & 65$\pm$5 & \cite{He_NatMat_2013} \\
ARPES & Gap-closing $T$ & 1UC FeSe/Nb:SrTiO$_3$ & 60$\pm$5 & \cite{Tan_NatMat_2013} \\
ARPES & Gap-closing $T$ & 1UC FeSe/Nb:SrTiO$_3$/KTaO$_3$ & 70 & \cite{Peng_PRL_2014} \\
ARPES & Gap-closing $T$ & 1UC FeSe/Nb:BaTiO$_3$/KTaO$_3$ & 75$\pm$2 & \cite{Peng_NatComm_2014}\\
ARPES & Gap-closing $T$ & 1UC FeSe/Nb:SrTiO$_3$ & 58$\pm$7 & \cite{Lee_Nat_2014} \\
4-probe & Zero resistance & 1UC FeSe/Nb:SrTiO$_3$ & 109 & \cite{Ge_NatMat_2015} \\
\hline\hline
\textbf{\textit{ex-situ}} & & & & \\
Transport & Zero resistance & Si/5UC FeSe/SrTiO$_3$ & $<$30 & \cite{Wang_CPL_2012} \\
& Onset $T$ & & 53 & \\
Transport & Zero resistance & Si/10UC FeTe/1UC FeSe/SrTiO$_3$ & 23.5 & \cite{Zhang_CPL_2014} \\
& Onset $T$ & & 40.2 & \\
Magnetization & Onset $T$ & Si/10UC FeTe/1UC FeSe/SrTiO$_3$ & 21 & \cite{Zhang_CPL_2014} \\
Magnetization & Onset $T$ & Si/10UC FeTe/3-4UC FeSe/SrTiO$_3$ & 20$-$45 & \cite{Deng_PRB_2014} \\
Magnetization & Onset $T$ & 10UC FeTe/1UC FeSe/Nb:SrTiO$_3$ & 85 & \cite{Sun_SciRep_2014} \\
Magnetization & Onset $T$ & Se/2UC FeSe/2UC Fe$_{0.96}$Co$_{0.04}$Se/ & 65 & \cite{Zhang_SciBull_2015} \\
& & 1UC FeSe/Nb:SrTiO$_3$ & & \\
\hline
\end{tabular}
\end{center}
\end{table}

A more robust proof of superconductivity would include (1) a zero-resistance state and (2) the Meissner effect (perfect diamagnetism). Due to air sensitivity, \textit{ex-situ} transport and thermodynamic measurements require film capping, with amorphous Se~\cite{Cui_PRL_2015}, amorphous Si~\cite{Wang_CPL_2012}, or epitaxial FeTe~\cite{Zhang_CPL_2014}. In all cases, film characteristics were degraded. Transport measurements of capped heterostructures have found a zero-resistance state below $\sim$20 K, and a rough onset temperature possibly up to $\sim$50 K. Similarly, magnetization measurements of capped samples have suffered from weak signals, broadened onset temperatures, or low superconducting volume fractions.

Given that many potential applications require some degree of atmosphere exposure, it remains crucial to investigate why capping, particularly epitaxial FeTe, has not worked well. FeTe possesses the same crystal structure as FeSe and its layers interact via van der Waals forces, so naively it should not create a severe disturbance of the FeSe layer below. Several hypotheses have been put forward. Ultrafast spectroscopy revealed an acoustic phonon mode in FeTe that may relax phonon-mediated pairing in FeSe~\cite{Tian_PRL_2016}. Alternatively, cross-sectional TEM revealed that intermixing with the capping layer can occur, whereby Te atoms substitute Se atoms in the FeSe monolayer~\cite{Li_PRB_2015}. As a third possibility, Zhao \textit{et al.} proposed that FeTe may hole-dope FeSe, reducing $T_c$~\cite{Zhao_BAPS_2016}.

In \textbf{Table~\ref{TabTc}}, we distinguish heterostructures that have conducting, Nb-doped SrTiO$_3$ from those that do not (undoped, bulk-insulating SrTiO$_3$). In general, transport measurements require an insulating SrTiO$_3$ substrate, but there are speculations that Nb-doped SrTiO$_3$ produces higher quality films. Sun \textit{et al.}~\cite{Sun_SciRep_2014} hinted that ``high quality FeSe films are easier to be achieved by MBE growth on conductive STO [SrTiO$_3$] substrates comparing to insulating STO substrates since the conductive STO substrate shows more flat and homogeneous surface for sample growth."

\subsubsection{\textit{In-situ}, micro-four-point measurements}

\begin{figure}[h]
\includegraphics{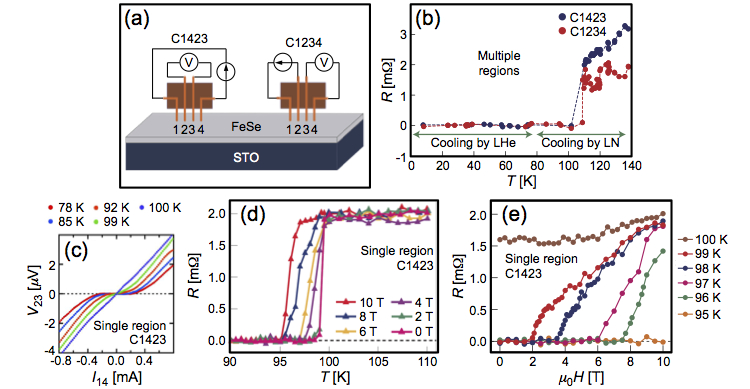}
\caption{(\textit{a}) Schematic of \textit{in-situ}, micro-four-point measurements of 1UC FeSe/SrTiO$_3$. Two possible configurations for applying a current and detecting a voltage drop are shown. (\textit{b}) Resistance vs. temperature plot, displaying a transition temperature $T_c$ = 109 K. The data points are extracted from $I$-$V$ curves acquired over different regions of the sample. (\textit{c}) Four-point $I$-$V$ curves, showing a metal-superconductor transition at a single location of the film. (\textit{d}) Resistance vs. temperature plot, with data points extracted from $I$-$V$ curves acquired at a fixed point on the sample with decreasing magnetic field, and increasing temperature at each field~\cite{Ge_private}. (\textit{e}) Magnetoresistance curves at various temperatures. Adapted from Ref.~\cite{Ge_NatMat_2015}.}
\label{FigJ}
\end{figure}

In late 2014, Ge \textit{et al.} reported an astonishing new record $T_c$ above 100 K in 1UC FeSe/SrTiO$_3$~\cite{Ge_NatMat_2015}. Here, we review their experiment in detail. The authors converted a commercial cryogenic STM into an \textit{in-situ}, micro-four-point probe by replacing the single STM tip with a set of four Cu/Au wires, separated by 10-100 $\mu$m (\textbf{Figure~\ref{FigJ}\textit{a}}). The four probes were collectively brought towards the sample at a 20$^{\circ}$ incline using the STM positioning system, until Ohmic contact with the sample was established for each probe. \textbf{Figure~\ref{FigJ}\textit{c}} shows several four-point $I$-$V$ curves, which transition from a nonlinear (superconducting, zero resistance) to linear (normal state, Ohmic) line shape as the temperature was raised above $T_c$.

Due to sample inhomogeneity or film damage from probes, linear $I$-$V$ curves were sometimes observed below $T_c$. As a result, Ge \textit{et al.} compiled resistance vs. temperature ($R$-$T$) plots in two manners. First, they acquired four-point $I$-$V$ measurements from separate locations for each temperature (\textbf{Figure~\ref{FigJ}\textit{b}}). As long as one $I$-$V$ curve per temperature showed signs of zero resistance, that temperature was deemed to be below $T_c$. With this method, Ge \textit{et al.} determined $T_c$ to be 109 K. Alternatively, they were also able to construct $R$-$T$ plots from measurements at one location, with a sequence of decreasing magnetic fields (\textbf{Figure~\ref{FigJ}\textit{d}}). With this second method, they demonstrated a similar $T_c$ of 99 K. The magnetoresistance measurements in \textbf{Figure~\ref{FigJ}\textit{e}} were also acquired at a fixed location.

We enumerate questions that have been raised about this experiment, and the authors' responses:
\begin{enumerate}
\item \textit{Question:} Is the result reproducible on multiple samples? \\ \textit{Response:} Ten different samples show similar results. Data from four samples are shown in the paper.
\item \textit{Question:} Is it possible that the authors simply lost a current lead contact as they cooled, resulting in a sudden drop of the measured $V$ to zero? \\ \textit{Response:} No, the authors measured full $I$-$V$ curves at each temperature and magnetic field ($B$), and extracted a $T$-dependent and $B$-dependent critical current (e.g. see \textbf{Figure~\ref{FigJ}\textit{d,e}}).
\item \textit{Question:} Is it possible that the actual $T$ of the sample is lower than the recorded $T$, giving the appearance of higher $T_c$?
\\ \textit{Answer:} No, careful calibration measurements show that the temperature of the sample is never more than 2 K less than the recorded temperature.
\item \textit{Question:} SrTiO$_3$ undergoes a structural transition at 105 K. Could this be responsible for the resistive transition observed at 109 K?\\ \textit{Response:} The authors performed a control experiment on bare, Nb-doped SrTiO$_3$, and showed that the structural transition produced a negligible signature in the $R$-$T$ plot [Figure 3\textit{b} of Ref.~\cite{Ge_NatMat_2015}]. 
\item \textit{Question:} Don't the measured values of $T_c$ = 109 K and $J_c$ = 1.3 $\times$ 10$^7$ A/cm$^{2}$ appear unexpectedly large?\\ \textit{Response:} The authors performed a control experiment on optimally-doped Bi$_2$Sr$_2$CaCu$_2$O$_{8+\delta}$, and found $T_c$ = 90 K, $J_c$ $\sim$ 6000 A/cm$^{2}$, in line with expectations. Their $J_c$ value is an order of magnitude higher than that of capped 1UC FeSe/SrTiO$_3$ films~\cite{Zhang_CPL_2014}, but similar to that of YBa$_2$Cu$_3$O$_{7-x}$ films~\cite{Zhu_SST_2013}.
\item \textit{Question:} Shouldn't there be a Berezinsky-Kosterlitz-Thouless (BKT) effect that broadens the resistive transition for a 2D superconductor? Why is the resistance drop so sharp (\textbf{Figure~\ref{FigJ}\textit{b}}), such that there are no data points within the transition~\cite{Bozovic_NatPhys_2014}? \\ \textit{Response:} Below $T_c$, conduction is 2D and restricted to the superconducting FeSe monolayer. Above $T_c$, conduction is shorted through the Nb-doped SrTiO$_3$ substrate, which is 3D and has a much lower resistivity than normal-state FeSe. Thus, the BKT transition is masked by shorting through the metallic substrate. It is also possible that there could be a proximity effect downward into SrTiO$_3$, such that the total system is not exactly 2D. The authors were able to collect data points within this sharp transition (\textbf{Figure~\ref{FigJ}\textit{d}}). 
\item \textit{Question:} In light of the previous question, why not use an insulating SrTiO$_3$ substrate? \\ \textit{Response:} The authors cited practical challenges~\cite{Ge_NatMat_2015}: ``Further limits exist for detecting films grown on an insulating substrate, as the feedback required to control the contact between the film and the tip is extremely difficult.''
\item \textit{Question:} Why doesn't the resistance change when the contact separation is increased tenfold~\cite{Bozovic_NatPhys_2014}? \\ \textit{Response:} When the probe separation distances are uniform, the resistance should scale with probe separation in both an infinite 2D conductor and a half-infinite 3D conductor. However, when the probe distances are unequal, their relationship to the overall resistance is more complicated [see Supplemental Information of Ref.~\cite{Ge_NatMat_2015}].
\item \textit{Question:} How could the resistive transition $T_c$ be higher than the gap-closing temperature $T_c$ measured \textit{in-situ} by ARPES? \\ \textit{Response:} ARPES averages signal over a beam spot size, but the \textit{in-situ} four-point probe may pick up filamentary superconductivity. Indeed, the authors found non-superconducting regions below $T_c$, but this could be attributed to both intrinsic sample inhomogeneity or film damage from probes. Alternatively, if the out-of-plane coherence length is short, superconductivity might be stronger at the bottom of the FeSe triple layer than at the top. ARPES and STM measure the top, but transport accesses the lowest-resistivity part, which may be located at the buried interface.
\item \textit{Question:} Is it possible that the apparent decreasing $T_c$ with increasing $B$ is simply due to gradual sample damage as $B$ is increased? \\ \textit{Response:} No, the authors showed the same result with increasing $B$ and decreasing $B$ at a fixed location.
\end{enumerate}

Despite intense scrutiny, we remain unaware of fatal flaws with the experiment by Ge \textit{et al.} Nevertheless, there are increasing calls for duplication of this result, as well as complementary \textit{in-situ} magnetization measurements of the Meissner effect~\cite{Bozovic_NatPhys_2014, Bozovic_NatPhys_2016}. The latter will require specific instrumentation, but will surely fill in an important piece of the puzzle.

\subsection{\label{growth}What are the necessary growth conditions?}

An accurate atomic structure is prerequisite to reliable modeling of electronic properties, and yet the former represents another significant experimental challenge for 1UC FeSe/SrTiO$_3$. Although SrTiO$_3$ is a workhorse substrate for MBE growth, it is notorious for its numerous nearly-degenerate surface reconstructions that sensitively depend on preparation conditions. With the (001) surface alone, O deficiency can drive the following reconstructions: 2$\times$1, 2$\times$2, $c$(4$\times$2), $c$(4$\times$4), 4$\times$4, $c$(6$\times$2), $\sqrt{5}$$\times$$\sqrt{5}$ $R$26.6$^{\circ}$, $\sqrt{13}$$\times$$\sqrt{13}$ $R$33.7$^{\circ}$~\cite{Lin_SS_2011}. Yet some feature of this complex surface interfaced with 1UC FeSe must generate a giant enhancement in $T_c$. Here, we examine and clarify growth procedures for 1UC FeSe/SrTiO$_3$. The overall challenge is to identify which steps are necessary and which are supplemental. In the following subsection, we review various measurements of the interface atomic structure.

\begin{figure}[h]
\includegraphics{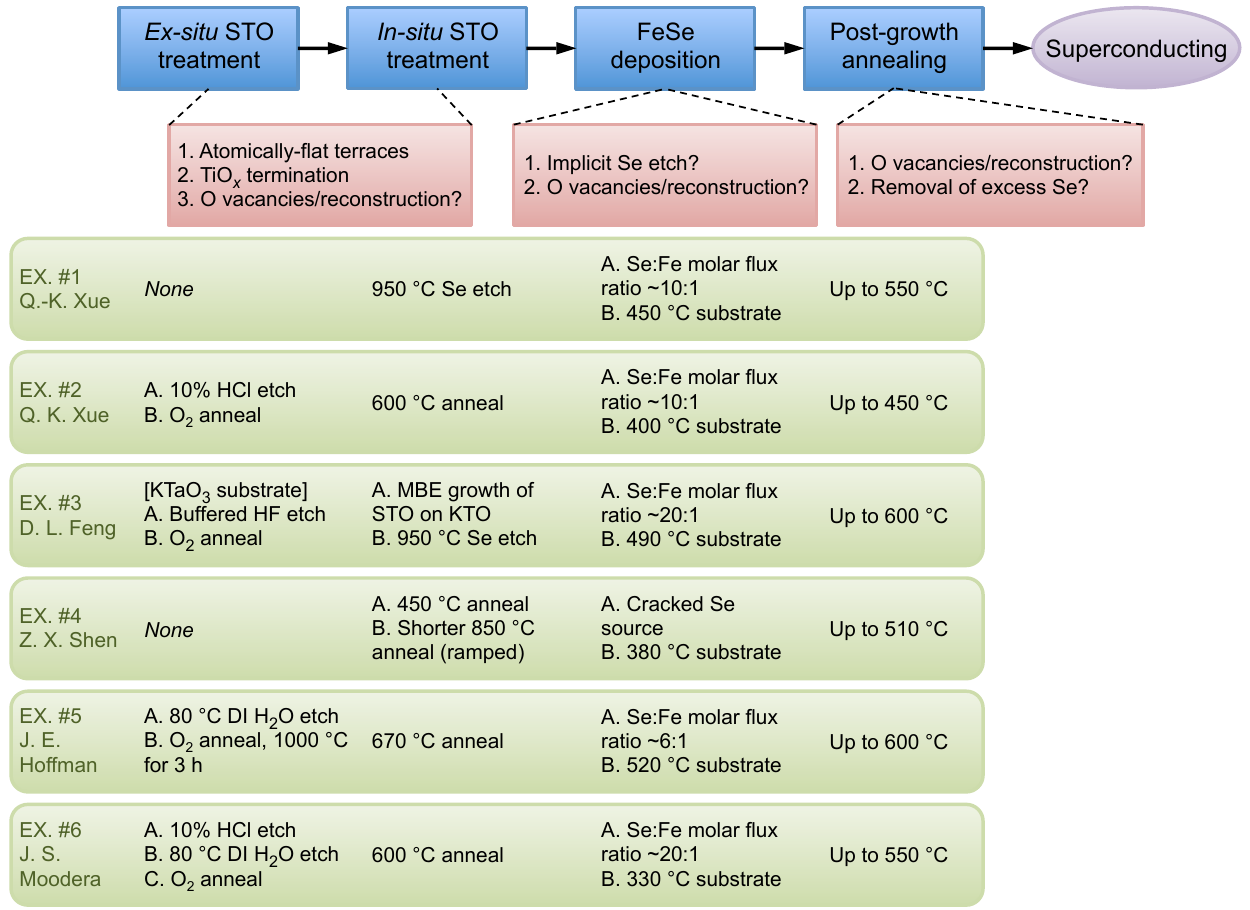}
\caption{Flowchart of growth procedure. The blue boxes highlight primary steps that lead to superconducting 1UC FeSe/SrTiO$_3$, while the red boxes describe the possible microscopic picture corresponding to each step. Six example recipes are given (green boxes): \#1 [Ref.~\cite{Wang_CPL_2012}], \#2 [Ref.~\cite{Zhang_CPL_2014}], \#3 [Refs.~\cite{Peng_PRL_2014, Tan_NatMat_2013}], \#4 [Ref.~\cite{Lee_Nat_2014}], \#5 [Ref.~\cite{Huang_PRL_2015}], and \#6 [Ref.~\cite{Zhao_BAPS_2016}].}
\label{FigFC}
\end{figure}

\textbf{Figure~\ref{FigFC}} presents a flowchart with typical growth recipes for 1UC FeSe/SrTiO$_3$. The recipes can be delineated into a few ``primary'' steps, which we discuss in turn. 

\subsubsection{SrTiO$_3$ treatment (\textrm{\textit{in situ}} and \textrm{\textit{ex situ}})}
Commerically available crystals of SrTiO$_3$ arrive with contaminated surfaces. In their original report, Wang \textit{et al.} introduced a novel strategy to clean Nb-doped SrTiO$_3$: they annealed the substrates in their MBE chamber at 950 $^{\circ}$C for 30 minutes, under a Se flux.~\cite{Wang_CPL_2012}. This treatment produced atomically-flat terraces amenable to STM imaging (albeit lacking atomic resolution). Subsequently, Bang \textit{et al.} hypothesized that this process created Se substitutions of surface O atoms~\cite{Bang_PRB_2013}. These Se$_{\scriptsize\textrm{O}}$ substitutions would then nucleate the growth of the first FeSe monolayer, leaving behind O vacancies that stabilize binding and donate electron carriers.  

Later films grown on insulating SrTiO$_3$ involved more conventional and better documented preparation protocols, involving an \textit{ex-situ} H$_2$O/acid etch followed by a high-temperature O$_2$ anneal in a tube furnace~\cite{Zhang_CPL_2014}. The H$_2$O/acid etch is believed to preferentially remove SrO, which has ionic bonding character, and leave behind a TiO$_2$-terminated surface~\cite{Kawasaki_Science_1994, Ohnishi_APL_2004, Connell_APL_2012}. It is thus unclear whether the previously-employed Se etch is a necessary procedure for growing epitaxial FeSe on SrTiO$_3$. Despite the explicit absence of this step here, it is possible that Se$_{\scriptsize\textrm{O}}$ substitutions are still generated during the deposition of 1UC FeSe.

\subsubsection{FeSe deposition}
To grow stoichiometric FeSe, two conditions are typically employed~\cite{Song_PRB_2011}. First, since Se is significantly more volatile than Fe, the substrate temperature is set between the source temperatures: $T_{\scriptsize\textrm{Fe}} > T_{\scriptsize\textrm{substrate}} > T_{\scriptsize\textrm{Se}}$. At least for growth on ``inert,'' graphitized SiC, this condition was rationalized as follows: Impinging Fe atoms with temperature $\sim$ $T_{\scriptsize\textrm{Fe}}$ will be adsorbed with sticking coefficient close to unity, while impinging Se atoms can stick only if they bind to free Fe on the substrate. Second, to compensate for high Se losses and to mitigate excess Fe clustering, typical molar flux ratios $\Phi_{\scriptsize\textrm{Se}}$/$\Phi_{\scriptsize\textrm{Fe}}$ range from 5 to 20. We note that with these two conditions (moderate substrate temperature 400-500 $^{\circ}$C, excess Se flux), there may still be a sizeable Se chemical potential at the SrTiO$_3$ surface driving the kinds of Se reactions proposed by Bang \textit{et al.}, but further investigations by STM or other techniques are needed.

\subsubsection{Post-growth annealing}
Post growth, the FeSe monolayer on SrTiO$_3$ becomes superconducting only after an additional vacuum anneal. He \textit{et al.} used ARPES measurements to show that in this process, the FeSe monolayer is progressively doped with electron carriers~\cite{He_NatMat_2013}. The electron doping induces a non-rigid band transformation that eventually leaves the FS with only electron pockets and opens up a gap. The source of electron doping remains an open question. He \textit{et al.} suggested that the electron doping could arise from O vacancies in SrTiO$_3$ created during annealing. Berlijn \textit{et al.} investigated the possibility of Se vacancies, but their calculations revealed Se vacancies to be hole dopants, not electron dopants~\cite{Berlijn_PRB_2014}. More recently, cross-sectional TEM imaging by Li \textit{et al.} suggested the presence of interstital Se atoms trapped at the FeSe/SrTiO$_3$ interface during growth, which are subsequently released upon annealing. The authors proposed that the removal of these interstitial Se atoms allows O vacancies in SrTiO$_3$ to effectively donate electron carriers to the FeSe monolayer~\cite{Li_2D_2016}.  

Overall, some elements of ``correct'' SrTiO$_3$ pre-treatment and post-growth annealing appear necessary to produce superconducting 1UC FeSe/SrTiO$_3$, but many aspects of the growth procedure could be clarified through more systematic investigations.

\subsection{What is the interface structure?}

We begin by comparing and contrasting three tools that have been applied to probe the interface atomic structure. 

\begin{enumerate}
\item Scanning tunneling microscopy (STM): real-space, atomic-resolution imaging of surface \\ \textit{Pro:} An \textit{in-situ} technique commonly integrated with a MBE chamber. \\ \textit{Con:} An indirect technique that requires additional modeling to make inferences about the buried interface.
\item Electron diffraction: low-energy (LEED) or reflection high-energy (RHEED): momentum-space information of surface atomic structure \\ \textit{Pro:} An \textit{in-situ} technique that can also monitor real-time growth (RHEED). \\ \textit{Con:} Phase information is unavailable. The interface signal may sometimes be buried after FeSe deposition~\cite{Lee_Nat_2014}.
\item Transmission electron microscopy (TEM): real-space, atomic-resolution imaging of exposed cross-section \\ \textit{Pro:} Direct atomic-resolution imaging of the interface cross-section. \\ \textit{Con:} An \textit{ex-situ} technique that requires capping (commonly FeTe). As evinced by Ref.~\cite{Li_PRB_2015, Li_2D_2016}, Te atoms from the cap may unintentionally intermix and substitute at least the top-layer Se atoms of 1UC FeSe. The size mismatch between Se and Te can strain the monolayer film, possibly altering its original binding structure to SrTiO$_3$. Additionally, the technique averages over each column of atoms in the $\sim$10-100 nm thick section being studied. 
\end{enumerate}

\subsubsection{2$\times$1 reconstruction}

\begin{figure}[h]
\includegraphics{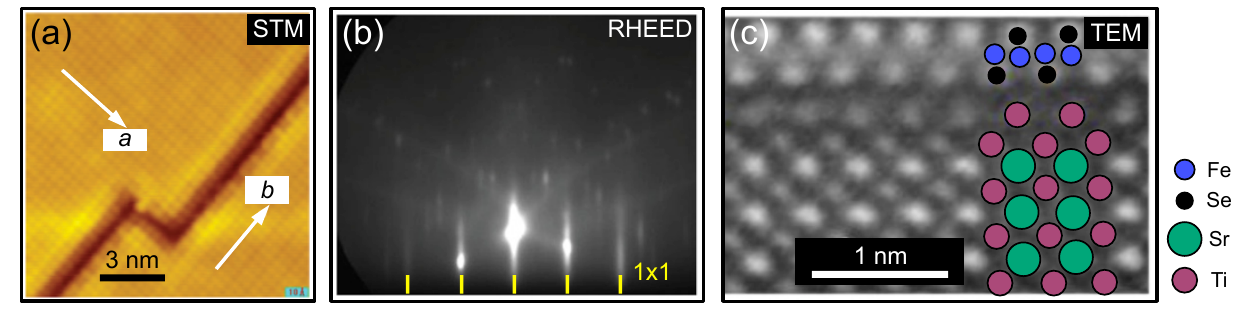}
\caption{Measurements of interface structure. (\textit{a}) STM topographic image showing orthogonal domains with dark stripes of 2$\times$1 periodicity. Across the trench, there is a half-UC phase shift. Adapted from Ref.~\cite{Bang_PRB_2013}. (\textit{b}) RHEED diffraction patterns of treated SrTiO$_3$ prior to FeSe deposition, exhibiting reconstruction spots. Adapted from Ref.~\cite{Lee_Nat_2014}. (\textit{c}) Cross-sectional TEM image of FeTe/1UC FeSe/SrTiO$_3$, revealing that the SrTiO$_3$ is terminated with a double-TiO$_x$ layer (atomic model is overlaid). Adapted from Ref.~\cite{Li_2D_2016}.}
\label{FigInt}
\end{figure}

The first hint of any interface superstructure was the appearance of dark stripes with 2$\times$1 periodicity in STM topographic images (\textbf{Figure~\ref{FigInt}\textit{a}})~\cite{Wang_CPL_2012, Bang_PRB_2013}. To explain this structure, Bang \textit{et al.} proposed an atomic model where half the O atoms on the surface TiO$_2$ layer are stripped off, and the bottom-layer Se atoms of the FeSe monolayer are laterally registered with the O vacancy sites~\cite{Bang_PRB_2013}. The authors argued that such arrangement could increase the binding energy, electron-dope the FeSe monolayer, and cause the FeSe monolayer to relax with a 2$\times$1 superstructure. In addition, since there are two equivalent O sites within a TiO$_2$ UC, the model could naturally explain the observation of half-UC phase shifts that occur either discontinuously at a trench~\cite{Li_JPCM_2014} (\textbf{Figure~\ref{FigInt}\textit{a}}), or continuously within a few nanometers of a domain boundary~\cite{Fan_NatPhys_2015}. 

As a word of caution, the 2$\times$1 stripes have not been universally observed. They are absent in AFM topographies~\cite{Li_APE_2013}, which might point to an electronic origin of the stripes, and are also absent in STM topographies of samples prepared in different ways~\cite{Zhang_PRB_2014, Huang_PRL_2015}. 

\subsubsection{More reconstructions}

Lee \textit{et al.} grew 1UC FeSe on SrTiO$_3$ with neither \textit{in-situ} Se etching nor \textit{ex-situ} treatment~\cite{Lee_Nat_2014}. They simply annealed as-bought substrates up to 830 $^{\circ}$C in their MBE chamber, until RHEED measurements detected superstructure spots, typically but not necessarily $\sqrt{5}$$\times$$\sqrt{5}$~\cite{Moore_BAPS_2015} (\textbf{Figure~\ref{FigInt}\textit{b}}). Subsequent deposition of FeSe and post-growth anneal resulted in superconducting samples with 1$\times$1 diffraction spots. 

Peng \textit{et al.} found a qualitatively different behavior in 1UC FeSe/BaTiO$_3$~\cite{Peng_NatComm_2014}. After annealing BaTiO$_3$ at 950 $^{\circ}$C under Se flux, their LEED images exhibited 3$\times$3 spots. Curiously, growth of 1UC FeSe produced three distinct domains: one domain commensurate with the BaTiO$_3$ 1$\times$1 UC, with expanded lattice constant 3.99 \AA; two domains rotated by $\pm$18.5$^{\circ}$, commensurate with a BaTiO$_3$ 3$\times$3 supercell, with smaller lattice constant 3.78 \AA. Furthermore, ARPES detected superconducting gaps in all three regions, with closing temperature $T_c$ ranging from 70-75 K. 

More recent experiments have detected superconducting gaps in 1UC FeSe on STO(110)~\cite{Zhou_APL_2016, Zhang_arXiv_2015(Ding)}, anatase TiO$_2$(001)~\cite{Ding_PRL_2016}, and rutile TiO$_2$~\cite{Rebec_arXiv_2016}, with different lattice constants and surface reconstructions (prior to growth). Taken together (\textbf{Table~\ref{Tab_Rec}}), the variety may imply that neither lattice constant nor the lateral atomic registry between 1UC FeSe and its underlying substrate are critical factors behind the enhanced superconductivity of this heterostructure.

\begin{table}[h]
\footnotesize
\caption{Reconstructions observed in various superconducting FeSe/($A$)TiO$_x$ heterostructures.}
\label{Tab_Rec}
\begin{center}
\begin{tabular}{c|c|c|c}
\hline
\cellcolor{shadecolor}\textbf{Substrate} & \cellcolor{shadecolor}\textbf{Lattice constant} &\cellcolor{shadecolor}\textbf{Reconstruction} & \cellcolor{shadecolor}\textbf{Ref.}\\
\cellcolor{shadecolor} & \cellcolor{shadecolor} [Bulk FeSe: $a_0$ = 3.77 \AA] & \cellcolor{shadecolor} & \cellcolor{shadecolor}\\ 
\hline
SrTiO$_3$(001) & $a_0$ = 3.90 \AA  & 2$\times$1 [STM] & \cite{Wang_CPL_2012, Bang_PRB_2013, Li_APE_2013, Li_JPCM_2014, Fan_NatPhys_2015} \\
& & $\sqrt{5}$$\times$$\sqrt{5}$ [RHEED] & \cite{Lee_Nat_2014, Moore_BAPS_2015}\\
& & $\sqrt{13}$$\times$$\sqrt{13}$ [various] & \cite{Zou_PRB_2016} \\
\hline
SrTiO$_3$(110) & $a_0$ = 3.90 \AA & 4$\times$1 [STM] & \cite{Zhou_APL_2016} \\
& $b_0$ = 5.52 \AA & 6$\times$1 [STM] & \cite{Zhou_APL_2016}\\
& & 3$\times$1 [LEED] & \cite{Zhang_arXiv_2015(Ding)} \\
\hline
BaTiO$_3$(001) & $a_0$ = 3.99 \AA & 3$\times$3 [LEED] & \cite{Peng_PRL_2014} \\
\hline
Anatase TiO$_2$(001) & $a_0$ = 3.78 \AA & 4$\times$1 [STM] & \cite{Ding_PRL_2016} \\
\hline
\end{tabular}
\end{center}
\end{table}

\subsubsection{Double-TiO$_x$ termination}

Perhaps what matters is the vertical structure of the interface. Using cross-sectional TEM, Li \textit{et al.}~\cite{Li_2D_2016} imaged a double-TiO$_x$ termination at the interface of FeTe/1UC FeSe/SrTiO$_3$ (\textbf{Figure~\ref{FigInt}\textit{c}}). Zou \textit{et al.}~\cite{Zou_PRB_2016} also uncovered a double-TiO$_x$ termination using x-ray diffraction, LEED and RHEED. Although such termination had long been proposed as a candidate model for the 2$\times$1 surface reconstruction~\cite{Erdman_Nat_2002}, it had largely been neglected in atomic models of 1UC FeSe/SrTiO$_3$ until this point. Roughly speaking, the extra TiO$_x$ termination is half as polar as a bulk TiO$_2$ layer, and helps SrTiO$_3$ mitigate a divergence of the electrostatic potential towards its bulk~\cite{Herger_PRL_2007}. Structural and ferroelectric properties are likely modified near this double-TiO$_x$ termination. Zou \textit{et al.} argued that the double-TiO$_x$ termination faciliates epitaxial growth of FeSe through stronger binding, and also improves charge transfer from oxygen vacancies~\cite{Zou_PRB_2016}.

Li \textit{et al.} also used TEM imaging to extract the structural parameters of their FeTe-capped sample. They found the 1UC FeSe to have a 9.5\% reduced chalcogen height with 2.5\% in-plane lattice tensile strain (compared to bulk values). Furthermore, within a $\sim$10 nm cross section, the authors imaged a lateral half-UC shift between the bottom Se atoms and topmost Ti atoms. If this feature is characteristic of uncapped 1UC FeSe/SrTiO$_3$, then it suggests that the heterostructure has local bond disorder due to lattice incommensuration. We note that STM $dI/dV$ measurements do reveal spectral and gap inhomogeneity even in pristine regions of FeSe with no in-plane, atomic-scale defects~\cite{Huang_PRB_2016}. Further systematic investigations and correlation of disorder with growth procedures is needed.

\section{ELECTRONIC STRUCTURE AND PAIRING}

Having surveyed a range of experiments characterizing the basic properties of 1UC FeSe/SrTiO$_3$, we turn to the question of electronic structure and pairing. Superconductors are typically categorized into one of two paradigms: conventional or unconventional (\textbf{Table~\ref{Tab_SC}}). In a conventional superconductor, electrons are bound into Cooper pairs by attractive interactions mediated by phonons. The resulting energy gap has $s$-wave angular symmetry and a uniform sign throughout the BZ. In an unconventional superconductor, many believe that quantum fluctuations from a proximate phase (e.g. magnetism) provide the glue to bind electrons. Since these fluctuations are often repulsive, the resulting gap function harbors sign changes throughout the BZ (to be further discussed in Subsection~\ref{subsec:MM}). This latter class of superconductors, which includes the cuprates and iron pnictides, has long been associated with higher $T_c$ values. However, the tables have turned with the recent discovery of 203 K conventional superconductivity in pressurized H$_3$S~\cite{Drozdov_Nat_2015}.

\begin{table}[h]
\footnotesize
\caption{Two paradigms of superconductivity.}
\label{Tab_SC}
\begin{center}
\begin{tabular}{c|c|c}
\hline
\cellcolor{shadecolor} & \cellcolor{shadecolor}\textbf{Conventional} & \cellcolor{shadecolor}\textbf{Unconventional} \\
\hline
\cellcolor{shadecolor} & 203 K & 164 K \\
\cellcolor{shadecolor}\textbf{Maximum $T_c$} & pressurized H$_3$S~\cite{Drozdov_Nat_2015} & pressurized HgBa$_2$Ca$_2$Cu$_3$O$_{8+\delta}$~\cite{Gao_PRB_1994} \\
\hline
\cellcolor{shadecolor}\textbf{Pairing mechanism} & Phononic & Electronic (magnetic/orbital)\\
\hline
\cellcolor{shadecolor}\textbf{Gap structure} & Sign-preserving & Sign-changing\\
\hline
\end{tabular}
\end{center}
\end{table}

\begin{figure}[h]
\includegraphics{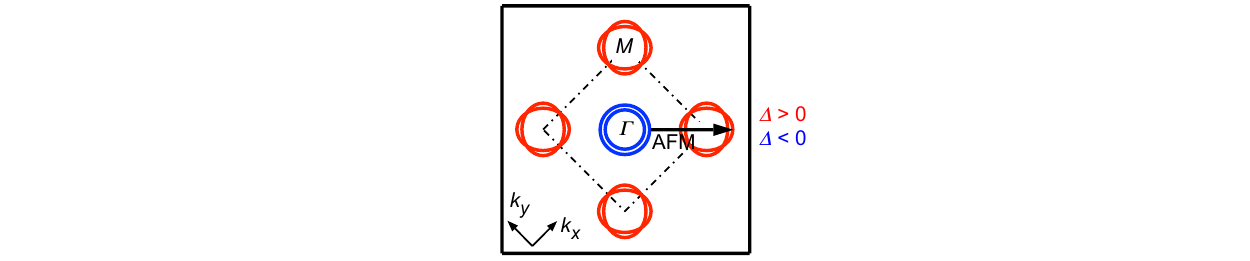}
\caption{Schematic of unconventional $s_{+-}$ pairing proposed for iron pnictide superconductors~\cite{Mazin_PRL_2008, Kuroki_PRL_2008}. The generic FS of these compounds consists of electron pockets at the BZ corner $M$ and hole pockets at the zone center $\Gamma$. The dashed line encloses the 2-Fe BZ. The electron and hole pockets are nested by an antiferromagnetic (AFM) wave vector, which can result in pairing if the gap function has one sign on the electron pockets (red, $\Delta > 0$) and the opposite sign on the hole pockets (blue, $\Delta < 0$).}
\label{Figs+-}
\end{figure}

Within months of the 2008 discovery of iron pnictide superconductors, Mazin \textit{et al.}~\cite{Mazin_PRL_2008} and Kuroki \textit{et al.}~\cite{Kuroki_PRL_2008} proposed an unconventional mechanism of pairing in these compounds. The basic premise was that first, the electron-phonon coupling constant was too small~\cite{Boeri_PRL_2008, Mazin_PRL_2008}; second, the proximity of the superconductor to an antiferromagnetic metal hinted at the role of spin fluctuations; and third, the multiband FS of these compounds, comprising electron pockets at the zone corner $M$ and hole pockets at the zone center $\Gamma$, could be crucial. The authors then argued that repulsive spin fluctuations, with wave vector spanning the separated electron and hole pockets, could pair electrons if the gap function reversed sign across the pockets with an overall ``$s_{+-}$'' symmetry (\textbf{Figure~\ref{Figs+-}}). Though not free from controversy~\cite{Onari_PRL_2009}, this framework prevailed in the early years of iron pnictide superconductors.

Fast-forwarding to the present, 1UC FeSe/SrTiO$_3$ poses several theoretical conundrums. First, its FS includes $M$ electron pockets (\textbf{Figure~\ref{FigFY}\textit{e}}), but not the $\Gamma$ hole pockets necessary for $s_{+-}$ pairing. Second, it appears to exhibit traits of both conventionality and unconventionality. In this section, we review contrasting indications for both phononic (Subsection~\ref{subsec:Phonons}) and electronic (spin/orbital, Subsection~\ref{subsec:Electronic}) mechanisms of pairing. We then evaluate ARPES and STM measurements of gap symmetry and structure (Subsection~\ref{subsec:Gapsym}). Finally, we discuss multiband, multiboson scenarios of pairing that enable phonons and spin/orbital fluctuations to operate constructively to enhance $T_c$ (Subsection~\ref{subsec:MM}). These latter ideas are far from being a \textit{fait accompli}, but exemplify a potential ``best-of-both-worlds'' path towards creating higher-$T_c$ superconductors.

\subsection{\label{subsec:Phonons}Phononic mechanisms}

In their original report, the discoverers of 1UC FeSe/SrTiO$_3$ proposed some sort of interface-enhanced electron-phonon coupling as the mechanism for high-$T_c$ superconductivity. Here, we discuss subsequent ARPES experiments by Lee, Schmitt, Moore \textit{et al.} that lent support to this notion~\cite{Lee_Nat_2014}.

\subsubsection{Replica bands}

\begin{figure}[h]
\includegraphics{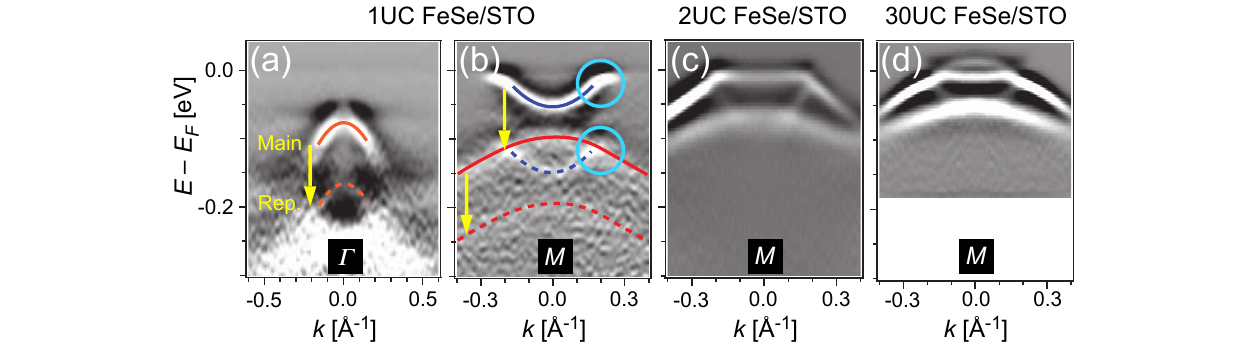}
\caption{(\textit{a}), (\textit{b}) ARPES high-symmetry cuts of 1UC FeSe/SrTiO$_3$ showing primary electronic bands (solid line overlays) and corresponding ``replica'' bands (dashed line overlays) -- fainter duplicates of the primary bands, shifted down by 100 meV (yellow arrows). These replica features suggest a $\vec{q}$ $\sim$ 0 coupling to a SrTiO$_3$ phonon mode. The blue circles highlight the duplication of a back-bending dispersion in the primary band as a superconducting gap opens near the Fermi energy. (\textit{c}), (\textit{d}) ARPES high-symmetry cuts of 2UC and 30UC FeSe/SrTiO$_3$, showing the absence of replica bands. Adapted from Ref.~\cite{Lee_Nat_2014}.}
\label{FigRB}
\end{figure}

What Lee, Schmitt, Moore \textit{et al.} discovered in their ARPES measurements was that each primary electronic band of 1UC FeSe/SrTiO$_3$ possessed a fainter replica band offset by 100 meV~\cite{Lee_Nat_2014}. These faint bands were near-duplicates of their primary counterparts, without being offset in momentum or smeared (\textbf{Figure~\ref{FigRB}\textit{a,b}}). In addition, the replica bands persisted at least to 120 K, well above the gap-opening temperature ($T_c$ = 58$\pm$7 K). Such replicas were absent in FeSe films two UC or thicker (\textbf{Figure~\ref{FigRB}\textit{c,d}}), pointing to an interfacial origin of these features. Similar phenomenology was observed by Peng \textit{et al.} in 1UC FeSe/BaTiO$_3$~\cite{Peng_NatComm_2014}.

In their interpretation of the replica bands, Lee, Schmitt, Moore \textit{et al.} first excluded the possibility of quantum-well states arising from 2D confinement. There is no reason for such states to have identical dispersions. Furthermore, quantum-well states exhibit a well-behaved dependence on layer, in contrast to the abrupt disappearance of replica bands in 2UC FeSe/SrTiO$_3$. Instead, the authors attributed the replica bands to bosonic shake-off, in analogy to vibrational shake-off observed in photoemission spectroscopy of H$_2$ molecules. They identified the boson with an optical O phonon band calculated for bulk SrTiO$_3$~\cite{Choudhury_PRB_2008}. Subsequent calculations of slab SrTiO$_3$ pointed to a surface phonon mode involving polar vibrations of vertical Ti-O bonds~\cite{Xie_SciRep_2015}. These theoretical comparisons were later corroborated by ARPES measurements that also found replica bands on bare SrTiO$_3$~\cite{Wang_NatMat_2016}. However, for an electron-phonon coupling $g(\vec{q})$ to produce nearly-identical bands with no momentum smearing, it must be sharply peaked at $\vec{q}$ = 0. This differs from the usual assumption of a constant $g(\vec{q})$ in theories of phonon-mediated superconductivity, and some modeling is needed to understand its origin.

\subsubsection{Model of interface electron-phonon coupling}

\begin{figure}[h]
\includegraphics{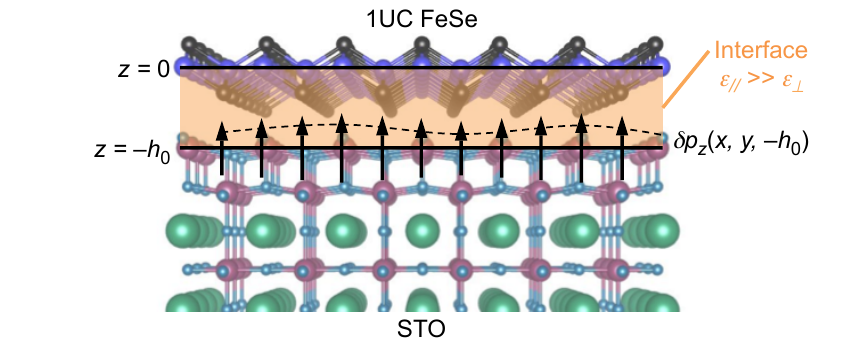}
\caption{Electrostatic model of interface electron-phonon coupling~\cite{Lee_Nat_2014, Lee_CPB_2015}, consisting of a 2D layer of FeSe at $z$ = 0 and a layer of vertical dipole moments $\delta p_z(x, y, -h_0)$ at the SrTiO$_3$ surface, $z$ = $-h_0$. If the interface region has anisotropic dielectric constants, $\epsilon_{\parallel}$ $\gg$ $\epsilon_{\perp}$, then the induced potential $\delta \Phi$ is exponentially peaked at $\vec{q}$ = 0.}
\label{FigTM}
\end{figure}

To explain how the electron-phonon coupling $g(\vec{q})$ could become sharply peaked at $\vec{q}$ = 0, Lee, Schmitt, Moore \textit{et al.} presented the following model~\cite{Lee_Nat_2014, Lee_CPB_2015}: Assume we have a 2D sheet of FeSe at $z$ = 0, and a layer of dipole moments below at the SrTiO$_3$ surface, $z = -h_0$ (\textbf{Figure~\ref{FigTM}}). The dipole moments come from vertical stretching of surface Ti-O bonds and are represented by $\delta p_z(x, y, -h_0)$. From an electrostatics calculation, these moments induce a potential at the FeSe layer,
\begin{equation}
\delta \Phi(x, y, 0) = \frac{\epsilon_{\parallel} h_0}{\epsilon^{3/2}_{\perp}} \int dx' dy' \frac{\delta p_z(x', y', -h_0)}{[\epsilon_{\parallel} h_0^2/\epsilon_{\perp} + (x-x')^2 + (y-y')^2]^{3/2}},
\end{equation}
where $\epsilon_{\parallel}$, $\epsilon_{\perp}$ are the in-plane and perpendicular dielectric constants in the interface region. Taking the Fourier transform yields
\begin{equation}
\delta \Phi(\vec{q}, 0) = \sqrt{\frac{\epsilon_{\parallel}}{\epsilon_{\perp}}} \frac{2\pi}{\sqrt{\epsilon_{\perp}}} \exp \bigg[-|\vec{q}| h_0 \sqrt{\epsilon_{\parallel}/\epsilon_{\perp}} \bigg] \delta p_z(\vec{q}, -h_0). 
\end{equation}
It follows that $g(\vec{q}) \propto \delta \Phi(\vec{q}, 0) \propto \exp (-|\vec{q}|/q_0)$, where $q^{-1}_0 = h_0 \sqrt{\epsilon_{\parallel}/\epsilon_{\perp}}$. Intuitively, the $\vec{q}$ $\sim$ 0 coupling hinges upon (1) the FeSe monolayer being sufficiently removed from the dipole layer (large $h_0$), and (2) the interface region screening lateral charge imbalance much more effectively than vertical charge imbalance (large $\epsilon_{\parallel}/\epsilon_{\perp}$).

Calculations by Rademaker \textit{et al.} showed that a ratio of $q_0/k_F$ $\sim$ 0.1 was needed for replica bands to duplicate primary band features without significant momentum smearing~\cite{Rademaker_NJP_2016}. If we take $k_F$ to be 0.20 \AA$^{-1}$~\cite{Liu_NatComm_2012} and $h_0$ to be 4.9 \AA~\cite{Li_2D_2016}, the distance between the surface TiO$_2$ layer and the Fe-plane, then $1/(h_0 k_F)$ $\sim$ 1 and we require $\epsilon_{\parallel}/\epsilon_{\perp}$ $\sim$ 100 in the interface region. While one should be wary of interpreting the model interface too literally, an argument suggests that it should contain contributions from both SrTiO$_3$ and FeSe, with the former having $\epsilon^{\scriptsize\textrm{STO}}_{\parallel}$ $\sim$ $\epsilon^{\scriptsize\textrm{STO}}_{\perp}$ in its 3D bulk limit, and the latter having $\epsilon^{\scriptsize\textrm{FeSe}}_{\parallel}$ $\gg$ $\epsilon^{\scriptsize\textrm{FeSe}}_{\perp}$ due to its 2D nature~\cite{Lee_CPB_2015}. 

Alternative speculations regarding the replica bands include O impurity bands~\cite{Mazin_NatMat_2015}, or some form of Raman scattering involving SrTiO$_3$ phonon modes~\cite{Kivelson_private}. No model details have been presented for impurity bands, and it is unclear whether sufficient cross section and viable selection rules exist for Raman scattering. Peaks and dips have been detected in STM filled- and empty-state $d^2I/dV^2$ spectra [Supplemental Material of Ref.~\cite{Huang_PRL_2015}], but the authors have not confirmed their identity as replica bands.  

\subsubsection{FeSe phonon modes}

In their initial STM measurements of 1UC FeSe/SrTiO$_3$, Wang \textit{et al.} reported two gaps in the $dI/dV$ point spectrum, at 9 meV and 20.1 meV respectively~\cite{Wang_CPL_2012}. This finding appeared to contradict early ARPES measurements of a single isotropic gap on the zone corner electron pockets, with $\Delta$ = 13$\pm$2 meV in one film and 15$\pm$2 meV in another film~\cite{Liu_NatComm_2012}. Coh \textit{et al.} offered an alternative explanation for the double-gap signature in terms of two FeSe phonon modes, which they argued could enhance $T_c$ when a monolayer of FeSe is locked to a SrTiO$_3$ substrate~\cite{Coh_NJP_2015}. Following this report, Tang \textit{et al.} examined $d^2I/dV^2$ point spectra of 1UC FeSe/SrTiO$_3$ and K-coated 2-4 UC FeSe/SrTiO$_3$~\cite{Tang_PRB_2016}. They identified positive-energy dips around 11 meV and 21 meV as FeSe phonon modes.

\subsection{\label{subsec:Electronic}Electronic mechanisms}

A feature of the interface phonon-coupling model is that it does not depend at all on 1UC FeSe, as long as the heterostructure has an interface dielectric constant that is sufficiently anisotropic. On one hand, such generality could be desirable for reproducing this mechanism in other systems. On the other hand, the model leaves open the possibility of preexisting pairing interactions within FeSe that are subsequently strengthened by SrTiO$_3$.

Two foil systems suggest that interface phonon coupling plays a secondary role to a primary pairing mechanism within FeSe that is enhanced by electron doping. The first is an FeSe-intercalate, (Li$_{1-x}$Fe$_x$)OHFeSe. Fe$_{\scriptsize\textrm{Li}}$ antisite substitutions increase electron transfer from buffer LiOH layers to FeSe~\cite{Chen_PRB_2016}, resulting in 40 K superconductivity~\cite{Lu_NatMat_2015}. Importantly, ARPES and STM measurements resolved low-energy bands that are nearly identical to those of 1UC FeSe/SrTiO$_3$ and gaps of similar magnitudes~\cite{Zhao_NatComm_2016, Niu_PRB_2015, Du_NatComm_2016, Yan_arXiv_2015}; however, no replica bands were visible. The second system involves coating FeSe with K adatoms, which inject electrons into the surface FeSe layer~\cite{Miyata_NatMat_2015, Wen_NatComm_2016, Ye_arXiv_2015, Tang_PRB_2015, Tang_PRB_2016, Zhang_NanoLett_2016, Song_PRL_2016}. The resulting superconducting phase has a gap-closing temperature up to 48 K, close to the 65 K value of 1UC FeSe/SrTiO$_3$. The electronic transition induced by progressive K deposition is rather rich and provides clues of unconventional mechanisms of pairing.

\subsubsection{Clues from the electron-doping phase diagram}

In the paradigm of unconventional superconductors, pairing is likely mediated by quantum fluctuations from nearby electronic phases. Hence, we glean inspiration from the electron-doping phase diagram of FeSe, keeping in mind that it may not be fully representative of 1UC FeSe/SrTiO$_3$.

\begin{figure}[h]
\includegraphics{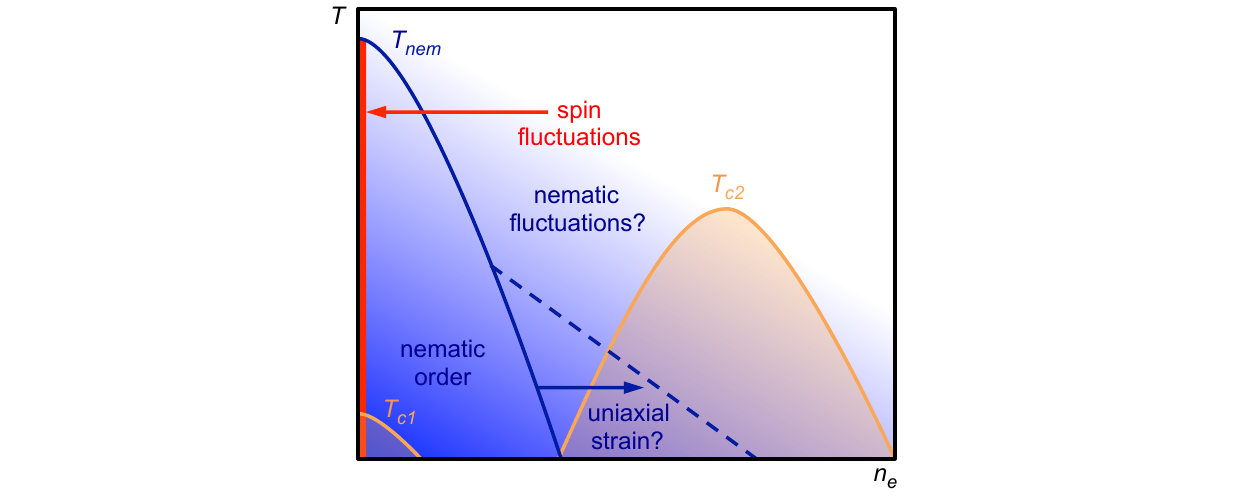}
\caption{Schematic phase diagram of electron-doped FeSe, consisting of two domes of superconductivity, and the possibility of nematic fluctuations. Adapted from Refs.~\cite{Miyata_NatMat_2015, Wen_arXiv_2015, Ye_arXiv_2015, Song_PRL_2016}. The existence of spin fluctuations in stoichiometric FeSe was reported by Refs.~\cite{Rahn_PRB_2015, Wang_NatMat_2015, Wang_arXiv_2015(2)}.}
\label{FigPD}
\end{figure}

The first striking feature in the electron-doping phase diagram of FeSe is that $T_c$ evolves through two domes~\cite{Song_PRL_2016}: a low-$T_c$ phase is first suppressed, eventually giving way to a higher-$T_c$ phase (\textbf{Figure~\ref{FigPD}}). In general, domes are hallmarks of unconventional superconductivity, less naturally explained within a purely phononic framework~\cite{Mazin_NatMat_2015}. Song \textit{et al.} also found that the higher-$T_c$ phase is insensitive to the disorder of nonmagnetic K adatoms, a point whose implications we revisit in Subsection~\ref{subsec:Gapsym}.

A second observation is that nematic order is suppressed preceding the high-$T_c$ phase~\cite{Miyata_NatMat_2015, Wen_NatComm_2016}, although a smaller overlapping tail of the nematic phase may persist due to remnant uniaxial strain from underlying bulk FeSe~\cite{Ye_arXiv_2015}. Nematic order is generally defined as broken rotational symmetry that preserves the translational symmetry of the crystal. In stoichiometric FeSe, nematic order is manifested as a small orthorhombic distortion~\cite{McQueen_PRL_2009} and a large splitting of the Fe $3d_{xz}$ and $3d_{yz}$ bands~\cite{Tan_NatMat_2013, Shimojima_PRB_2014, Nakayama_PRL_2014, Watson_PRB_2015, Zhang_PRB_2015, Zhang_arXiv_2015}, without concomitant magnetic order~\cite{Medvedev_NatMat_2009}. Given the proximity and possible overlap of the nematic phase, it is tempting to ask whether nematic quantum criticality could be at play. Nematic fluctuations would provide attractive $\vec{q} \sim 0$ interactions that help bind electrons~\cite{Fernandes_SST_2012, Yamase_PRB_2013, Lederer_PRL_2015, Dumitrescu_arXiv_2015, Li_SciBull_2016, Kang_arXiv_2016}, much like the aforementioned SrTiO$_3$ phonons.

\subsubsection{Nematic fluctuations}

Since 1UC FeSe bound to SrTiO$_3$ is nominally tetragonal, nematic order should be globally suppressed. However, if there truly exists a large underlying nematic susceptibility that produces fluctuations, then nanoscale patches of such fluctuations might be pinned around crystalline imperfections that locally break tetragonal symmetry.

\begin{figure}[h]
\includegraphics{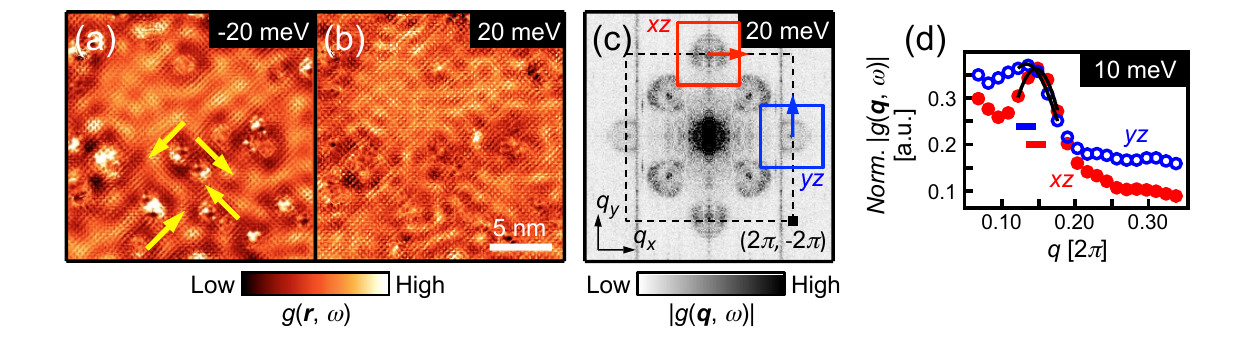}
\caption{Bounds on nanoscale nematicity in 1UC FeSe/SrTiO$_3$. (\textit{a}), (\textit{b}) STM conductance maps $g(\vec{r}, \omega)$ over a region of 1UC FeSe/SrTiO$_3$ containing several atomic-scale defects, revealing dispersive quasiparticle interference (QPI) patterns. The prevalent defects are anisotropic and appear in four possible orientations (yellow arrows). (\textit{c}) Fourier transform amplitude $|g(\vec{q}, \omega)|$ of (\textit{b}). The red and blue boxes enclose ring intensities that arise from scattering between Fermi electron pocket states of Fe $3d_{xz}$ and $3d_{yz}$ orbital characters, respectively. (\textit{d}) Normalized line cuts across the arrows in (\textit{c}), used to compare $xz$/$yz$ scattering wave vectors. The horizontal bars mark the peak locations determined from Gaussian fits (solid lines), with inherent resolution $\delta q$ = 0.028 $\pi$. No signature of orbital nematicity was detected. Adapted from Ref.~\cite{Huang_PRB_2016}.}
\label{FigQPI}
\end{figure}

Using STM as a nanoscale probe, Huang \textit{et al.} investigated quasiparticle interference (QPI) patterns generated around anistropic defects in 1UC FeSe/SrTiO$_3$~\cite{Huang_PRB_2016} (\textbf{Figure~\ref{FigQPI}}). Since QPI anisotropy can arise from random disorder or experimental artifacts, the authors developed a realistic, $T$-matrix model to specifically detect orbital anisotropy of Fe $3d_{xz}$ and $3d_{yz}$ bands. By sampling multiple spatial regions of a film, they excluded $xz/yz$ orbital ordering with domain size larger than $\delta r^2$ = 20 nm $\times$ 20 nm, $xz/yz$ Fermi wave vector difference larger than $\delta k$ = 0.014 $\pi$, and energy splitting larger than $\delta E$ = 3.5 meV. The lack of detectable ordering pinned around defects disfavors scenarios of a proximate nematic quantum critical point in 1UC FeSe/SrTiO$_3$, in contrast to K-coated FeSe (\textbf{Figure~\ref{FigPD}}).

\subsubsection{Spin fluctuations}

Given the importance of spin fluctuations in many iron pnictide superconductors, their role in pairing should also be considered in 1UC FeSe/SrTiO$_3$~\cite{Linscheid_arXiv_2016}. Several inelastic neutron scattering measurements found that in stoichiometric FeSe, there are stripe spin fluctuations that are enhanced below the orthorhombic transition temperature~\cite{Rahn_PRB_2015, Wang_NatMat_2015, Wang_arXiv_2015(2)} (\textbf{Figure~\ref{FigPD}}). However, magnetic order is absent, owing to some sort of frustration~\cite{Glasbrenner_NatPhys_2015, Wang_NatPhys_2015, Chubukov_PRB_2015} or quadrupolar order~\cite{Yu_PRL_2015}. The nature of spin excitations in 1UC FeSe/SrTiO$_3$ remains an important open question, especially since both doping and the Se height may tune exchange interactions. Some experimental ingenuity is required, as \textit{ex-situ} neutron scattering measurements are likely not feasible on 1UC films. One possible approach is the use of STM to map the magnetic-field dependence of impurity signatures, in comparison with theoretical modeling~\cite{Gastiasoro_arXiv_2016}.

\subsection{\label{subsec:Gapsym}Gap symmetry and structure}

We shift gears and consider pairing from the viewpoint of gap symmetry and structure. In general, such questions have proven more challenging to address in the iron-based superconductors than in the cuprates. Unlike the cuprates, with a single Cu $d$ band and universal $d_{x^2-y^2}$ gap symmetry, the multiband FS the of iron-based superconductors can allow a variety of gap structures across their member compounds. Even within the same compound, such as KFe$_2$As$_2$, different pairing symmetries can be tuned by pressure~\cite{Tafti_NatPhys_2013}. To add to the challenge, many candidate gap structures share the same angular symmetry~\cite{Hirschfeld_RPP_2011}, and thus cannot be differentiated by the corner junction experiments that proved instrumental in revealing the $d$-wave gap of YBa$_2$Cu$_3$O$_{7-x}$~\cite{Wollman_PRL_1993}. ARPES can resolve gap magnitudes on each specific band, but not their signs. STM QPI measurements carry phase-sensitive information, but can be challenging to interpret or normalize~\cite{Hirschfeld_PRB_2015}. 

With these complications in mind, there is less likely to be a clear,``smoking-gun'' experiment revealing the gap symmetry of 1UC FeSe/SrTiO$_3$. A more likely scenario is that through multiple experimental measurements, consensus will begin to converge upon a candidate gap function.

\subsubsection{The candidates}

\begin{figure}[h]
\includegraphics{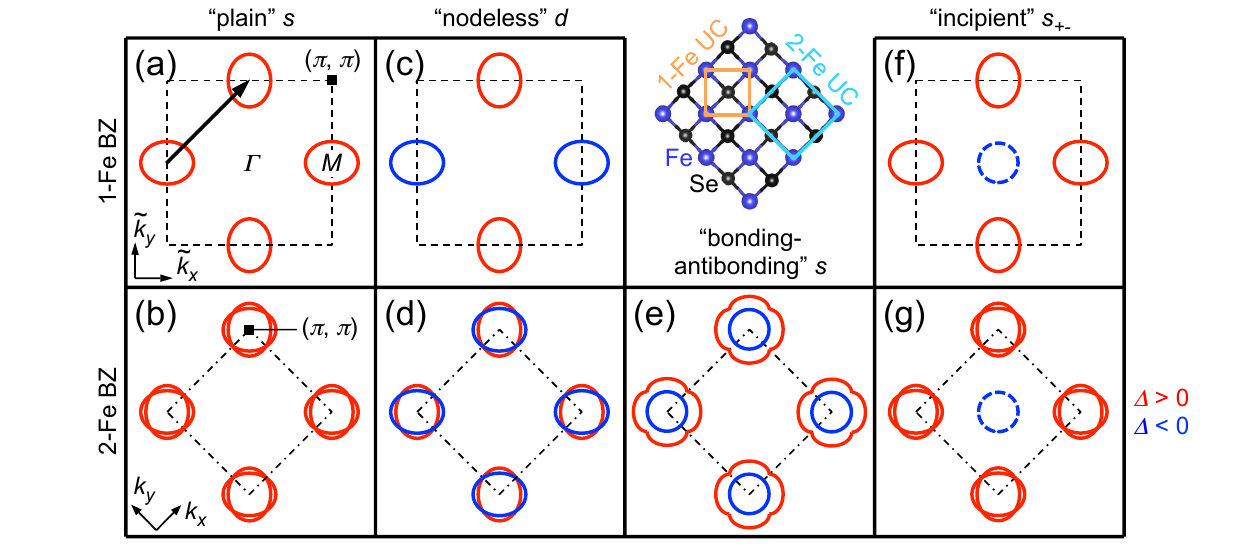}
\caption{Gap symmetry candidates for 1UC FeSe/SrTiO$_3$. The top row depicts these structures in a 1-Fe Brillouin zone (BZ) ($\tilde{\vec{k}}$), and the bottom row depicts their folded counterparts in the 2-Fe BZ ($\vec{k}$). The arrow in (a) marks the folding wave vector, and the schematic illustrates the 1-Fe and 2-Fe unit cells (UCs). Note that the ``bonding-antibonding'' $s$ structure in (\textit{e}) cannot be unfolded. The dashed pockets for ``incipient'' $s_{+-}$ in (\textit{f}), (\textit{g}), represent bands that lie completely below (or above) the Fermi energy.}
\label{FigC}
\end{figure}

Given the Fermi surface of 1UC FeSe/SrTiO$_3$, with only electron pockets, the primary gap symmetry candidates are ``plain'' $s$, ``nodeless'' $d$, ``bonding-antibonding'' $s$, and ``incipient'' $s_{+-}$ (\textbf{Figure~\ref{FigC}}). Nodal candidates are inconsistent with the fully-gapped structures detected by STM and ARPES (\textbf{Figure~\ref{FigFY}\textit{b}}).

``Plain'' $s$ gap symmetry (\textbf{Figure~\ref{FigC}\textit{a,b}}) will be discussed in Subsection~\ref{subsec:Fan}. ``Nodeless'' $d$ (\textbf{Figure~\ref{FigC}\textit{c,d}}), which appears most similar to the gap in cuprates, is strictly defined in a 1-Fe, ``pseudocrystal momentum'' BZ ($\tilde{\vec{k}}$) that only exists when FeSe has exact glide-plane symmetry. When the gap structure is folded into the proper 2-Fe BZ, it is no longer $d$-wave with respect to regular crystal momentum ($\vec{k}$). In addition, nodes are technically created when opposite-sign gaps meet at the pocket crossings~\cite{Mazin_PRB_2011}. However, based on microscopic details, the nodal quasiparticle weight could be weak and elude spectroscopic detection~\cite{Maier_PRB_2011, Kreisel_PRB_2013}. 

If the folded pockets in the 2-Fe BZ hybridize and detach from each other, then nodes will certainly be avoided, leading to a ``bonding-antibonding'' $s$ scenario (\textbf{Figure~\ref{FigC}\textit{e}}). Here, the inner and outer pockets host gaps of opposite sign. 

``Incipient'' $s_{+-}$ posits that an opposite-sign gap develops on a sunken zone center hole pocket (\textbf{Figure~\ref{FigC}\textit{f,g}}).
In the weak-coupling limit, such a gap can still be sizeable~\cite{Chen_PRB_2015}, the reasons for which we discuss in Subsection~\ref{subsec:MM}.

\subsubsection{ARPES measurements}

\begin{figure}[h]
\includegraphics{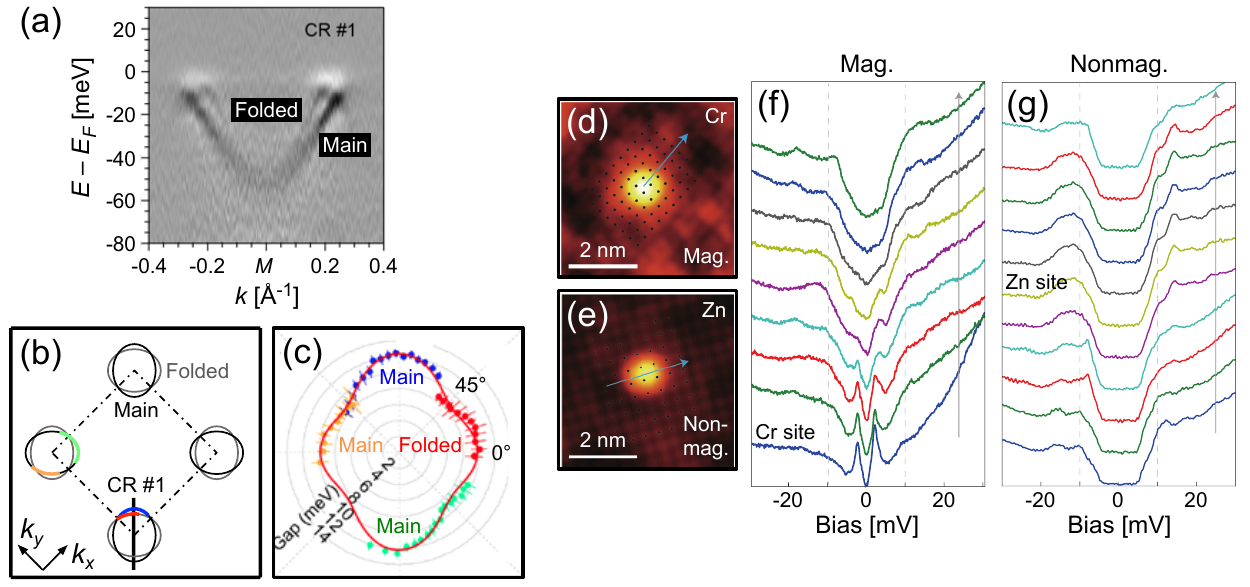}
\caption{(\textit{a})-(\textit{c}) ARPES measurements of gap anisotropy. The nearly-identical gap structures on equivalent segments of the main and folded pockets (red and orange cuts) disfavors scenarios of ``bonding-antibonding'' $s$-wave pairing. Adapted from Ref.~\cite{Zhang_arXiv_2015(2)}. (\textit{d})-(\textit{g}) STM $dI/dV$ line cuts across magnetic (Cr) and non-magnetic (Zn) adatoms. In-gap states are produced in the former, while no changes are visible in the latter. These observations favor ``plain'' $s$-wave symmetry. Adapted from Ref.~\cite{Fan_NatPhys_2015}.}
\label{FigGS}
\end{figure}

Early ARPES investigations of 1UC FeSe/SrTiO$_3$ reported isotropic gaps ($\Delta$ = 13-15 meV) on nearly-circular electron pockets~\cite{Liu_NatComm_2012, He_NatMat_2013, Tan_NatMat_2013, Lee_Nat_2014}. To resolve finer structure, Peng \textit{et al.} grew 1UC FeSe on SrTiO$_3$/KTaO$_3$, whose expanded lattice constant increased pocket ellipticity~\cite{Peng_PRL_2014}. Alternatively, Zhang \textit{et al.} changed photon polarizations to selectively probe bands of different orbital characters~\cite{Zhang_arXiv_2015(2)}. In both cases, the authors observed two pockets at each corner (main and folded), with no signs of hybridization (\textbf{Figure~\ref{FigGS}\textit{a-c}}). Momentum distribution cuts across the intersection of the main and folded pockets revealed a single band, with no detectable splitting. Furthermore, gap measurements on equivalent segments of the main and folded pockets showed identical structure. Such lack of sizeable hybridization remains to be understood, given that both spin-orbit coupling or the SrTiO$_3$ substrate can break glide-plane symmetry. More importantly, it also disfavors scenarios of ``bonding-antibonding'' $s$-wave pairing. The authors in both reports also resolved gap anisotropy, with minima directed along the Fe-Se axes. These measurements will provide useful feedback for theoretical gap function calculations.

\subsubsection{\label{subsec:Fan}STM measurements}

Fan \textit{et al.} employed a multi-pronged STM approach, involving phase-sensitive QPI and defect imaging, to build support for ``plain'' $s$-wave superconductivity in 1UC FeSe/SrTiO$_3$~\cite{Fan_NatPhys_2015}. In particular, the authors found that magnetic adatoms (Cr, Mn) induced in-gap bound states, while non-magnetic adatoms did not (Zn, Ag, K) (\textbf{Figure~\ref{FigGS}\textit{d-g}}). This observation is consistent with an underlying gap structure without sign changes, but not a fool-proof guarantee of such. Anderson's theorem states that a superconductor with a sign-preserving gap should be robust against the disorder of nonmagnetic impurities. Taken in its equivalent, contrapositive form, the observation of in-gap states induced by nonmagnetic impurities would thereby signal a sign-changing gap. However, the converse statement (``robustness against nonmagnetic impurities'' $\Longrightarrow$ ``sign-preserving gap'') is not logically identical to the original theorem, so it lacks a``smoking-gun'' nature~\cite{Hirschfeld_RPP_2011, Beaird_PRB_2012}. In the case of 1UC FeSe/SrTiO$_3$, with the ``nodeless'' $d$ and ``bonding-antibonding'' $s$ gap structures, the opposite-sign gaps reside on normal-state Fermi pockets with different orbital characters. Given that the impurities in the Fan \textit{et al.} experiment outside the Fe-plane, they may have had insufficient interorbital scattering strength to produce a pair-breaking effect. This complication is alleviated in the case of $s_{+-}$ pairing in the iron pnictides. Since both the electron and hole pockets hosting opposite-sign gaps share the same orbital characters, interband scattering mixing the signs is easier.

Despite these caveats, the preponderance of current experiments favor same-sign gaps on all Fermi pockets of 1UC FeSe/SrTiO$_3$ above the other possibilities.

\subsection{\label{subsec:MM}Multiband, multiboson scenarios of pairing}

In this final subsection, we examine pairing scenarios in which multiple bosons work cooperatively across multiple bands to boost $T_c$. More specifically, we consider various ways in which attractive interactions (e.g., mediated by phonons) and repulsive interactions (e.g., mediated by spin fluctuations) can fit under the same roof in 1UC FeSe/SrTiO$_3$~\cite{Xiang_PRB_2012, Lee_Nat_2014, Li_SciBull_2016, Chen_PRB_2015}.

The basic picture can be explained from the $T=0$ gap equation of a one-band superconductor in the weak-coupling limit: 
\begin{equation}
\label{EqBCS}
\Delta_{\vec{k}} = -\sum_{\vec{k}'}\frac{V_{\vec{k}, \vec{k}'} \Delta_{\vec{k}'}}{2E_{\vec{k'}}}. 
\end{equation}
Here, $V_{\vec{k}, \vec{k}'}$ is an effective potential that scatters a Cooper pair from ($\vec{k}$$\uparrow$, $-\vec{k}$$\downarrow$) to ($\vec{k}'$$\uparrow$, $-\vec{k}'$$\downarrow$), $E_{\vec{k'}} = \sqrt{\xi^2_{\vec{k'}} + |\Delta^2_{\vec{k'}}|} > 0$ is the Bogoliubov quasiparticle energy, $\xi_{\vec{k'}}$ is the normal-state quasiparticle energy, and $\Delta_{\vec{k}}$ is the gap function. Since any such $\Delta_{\vec{k}}$ must obey Eq.~\ref{EqBCS} self-consistently, its form is determined by $V_{\vec{k}, \vec{k}'}$ as follows: 
\begin{enumerate}
\item Attractive interactions ($V_{\vec{k}, \vec{k}'} < 0$) increase the gap amplitude if they connect segments of the FS hosting same-sign gaps ($\Delta_{\vec{k}} > 0$ and $\Delta_{\vec{k'}} > 0$, or $\Delta_{\vec{k}} < 0$ and $\Delta_{\vec{k'}} < 0$).
\item Repulsive interactions ($V_{\vec{k}, \vec{k}'} > 0$) increase the gap amplitude if they connect segments of the FS hosting opposite-sign gaps ($\Delta_{\vec{k}} > 0$ and $\Delta_{\vec{k'}} < 0$, or $\Delta_{\vec{k}} < 0$ and $\Delta_{\vec{k'}} > 0$).
\end{enumerate}
Attractive and repulsive interactions can therefore simultaneously increase the gap amplitude, if the interactions connect \textit{different} segments of the FS, with appropriate signs in the gap function. In the limit of forward scattering ($\vec{k} = \vec{k}'$), attractive interactions have the form $V_{\vec{k}, \vec{k}'} \propto -\delta_{\vec{k}, \vec{k'}}$, and from Eq.~\ref{EqBCS}, increase the gap amplitude irrespective of the gap sign or functional form. 

\begin{figure}[h]
\includegraphics{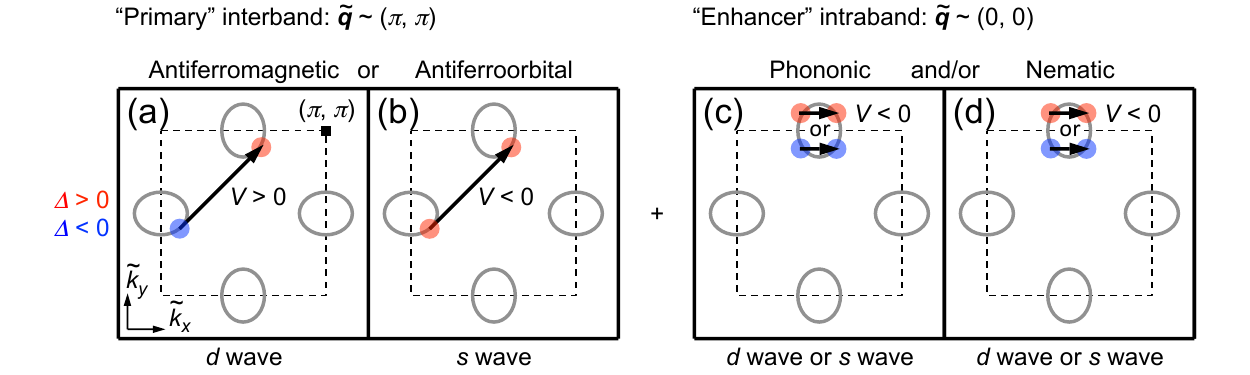}
\caption{Multiband, multiboson scenarios of pairing in 1UC FeSe/SrTiO$_3$. For simplicity, we work in a 1-Fe, pseudocrystal momentum ($\tilde{\vec{k}}$) BZ. (\textit{a}), (\textit{b}) \textit{Step 1} - the gap ($\Delta$) symmetry is determined by a ``primary'' interband interaction, peaked around $\tilde{\vec{q}} = (\pi, \pi)$, which may be attractive ($V< 0$) or repulsive ($V > 0$). (\textit{c}), (\textit{d}) \textit{Step 2} - $T_c$ is further boosted by ``enhancer" intraband interactions that are attractive. Due to their forward-scattering nature [i.e., peaked around $\tilde{\vec{q}} = (0,  0)$], they can raise $T_c$ for any gap symmetry.}
\label{FigMB}
\end{figure}

\textbf{Figure~\ref{FigMB}} illustrates a pairing framework for 1UC FeSe/SrTiO$_3$ involving multiple bosons. We postulate the existence of a ``primary'' interband interaction peaked around $\tilde{\vec{q}} = (\pi, \pi)$, connecting the disparate electron pockets and dictating the overall gap symmetry (\textbf{Figure~\ref{FigMB}\textit{a,b}}). This interaction could be a repulsive antiferromagnetic spin fluctuation, stabilizing $d$-wave pairing, or an attractive antiferroorbital fluctuation, stabilizing $s$-wave pairing. Then in addition, there may be ``enhancer'' intraband interactions that are necessarily attractive and peaked around $\tilde{\vec{q}} = (0, 0)$ (\textbf{Figure~\ref{FigMB}\textit{c,d}}). These interactions universally boost pairing irrespective of the underlying gap symmetry, and may come from SrTiO$_3$ phonons~\cite{Lee_Nat_2014}, and/or nematic fluctuations.    

Although this pairing framework is appealing due to its inclusive nature, we emphasize that other than the SrTiO$_3$ phonon mode, there have been no experimental indications of the other interactions shown in \textbf{Figure~\ref{FigMB}}. Some suggest that DFT calculations of 1UC FeSe/SrTiO$_3$ with checkerboard antiferromagnetism [$\tilde{\vec{q}} = (\pi, \pi)$] best resemble experimental data~\cite{Bazhirov_JPCM_2013, Zheng_SciRep_2013, Coh_NJP_2015}, thus motivating the possible existence of related spin fluctuations. Others take the orbitally-ordered state of bulk FeSe as a hint of possible ferroorbital [nematic, $\tilde{\vec{q}} = (0, 0)$] or antiferroorbital [$\tilde{\vec{q}} = (\pi, \pi)$] fluctuations~\cite{Dumitrescu_arXiv_2015, Li_SciBull_2016}.  

\subsubsection{Pairing involving incipient bands}

Alternatively, we recall that inelastic neutron scattering measurements have detected stripe spin fluctuations [$\tilde{\vec{q}} = (\pi, 0)$] in bulk FeSe~\cite{Rahn_PRB_2015, Wang_NatMat_2015, Wang_arXiv_2015(2)}, similar to many iron pnictide compounds. At first glance, it is unclear whether such interactions, if they persist in 1UC FeSe/SrTiO$_3$, would be useful for pairing. The usual hole pocket at the BZ center, located $\Delta \tilde{\vec{k}} = (\pi, 0)$ away from the Fermi electron pockets, is sunken 65-80 meV below the Fermi energy (\textbf{Figure~\ref{FigFY}}). However, ARPES measurements have demonstrated that in LiFeAs, a superconducting gap can develop on a sunken hole pocket 10 meV below the Fermi energy~\cite{Miao_NatComm_2015}. Motivated by this observation, Chen \textit{et al.} proposed a ``bootstrap'' mechanism of pairing involving incipient bands in 1UC FeSe/SrTiO$_3$~\cite{Chen_PRB_2015}.

\begin{figure}[h]
\includegraphics{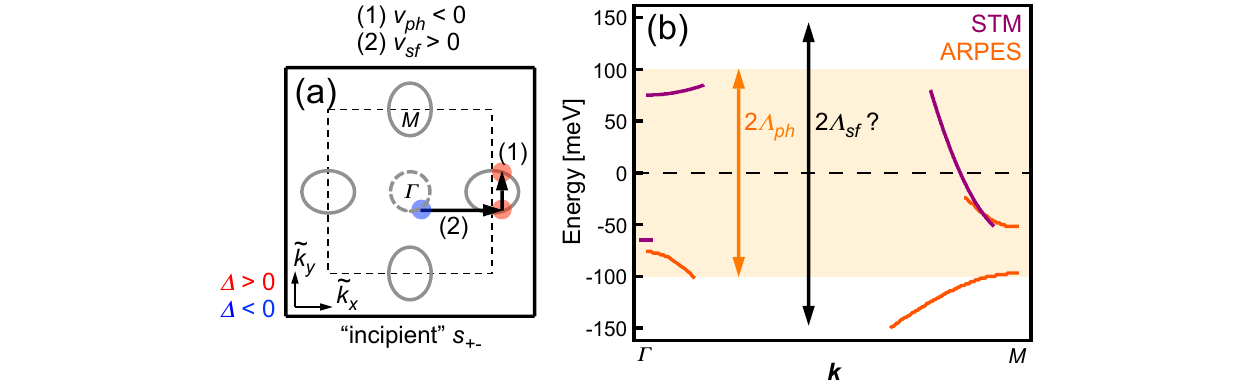}
\caption{``Bootstrap'' mechanism of pairing~\cite{Chen_PRB_2015}. (\textit{a}) \textit{Step 1} - An attractive intraband interaction, such as $\tilde{\vec{q}} \sim (0, 0)$ coupling to SrTiO$_3$ phonons, opens up a same-sign gap within the Fermi pockets at $M$. \textit{Step 2} - A repulsive interband interaction, possibly $\tilde{\vec{q}} \sim (\pi, 0)$ spin fluctuations, ``bootstraps'' to the $M$ pocket gaps and opens up an additional opposite-sign gap on the sunken hole pocket at $\Gamma$. Together, the gap function has overall $s_{+-}$ symmetry. (\textit{b}) Band structure diagram informed by STM~\cite{Huang_PRL_2015} and ARPES~\cite{Lee_Nat_2014} measurements. The energy scales of phonons and spin fluctuations (interactions $v_{ph}$, $v_{sf}$ and cutoffs $\Lambda_{ph}$, $\Lambda_{sf}$), along with the low-lying bands, determine the extent to which ``bootstrapping'' can enhance $T_c$.}
\label{FigBS}
\end{figure}

The basic picture proposed by Chen \textit{et al.} is illustrated in \textbf{Figure~\ref{FigBS}\textit{a}}. In the weak-coupling limit, interactions such as $\tilde{\vec{q}} = (\pi, 0)$ spin fluctuations between a Fermi sheet and an incipient band cannot open up a superconducting gap by themselves. However, if there are preexisting interactions, such as phonons, that stabilize a small gap within the Fermi sheet, then spin fluctuations from incipient bands can come into play and enhance pairing. In 1UC FeSe/SrTiO$_3$, this mechanism results in ``incipient'' $s_{+-}$ symmetry, where the Fermi electron pockets host same-sign gaps and the sunken hole pocket at the zone center hosts an opposite-sign gap. From the author's calculations, incipient bands can boost $T_c$ by an order of magnitude, although numbers are sensitive to estimates of interaction strengths and cutoff. 

An appealing feature of this model in the context of 1UC FeSe/SrTiO$_3$ is that the Fermi electron pockets and sunken hole pocket barely or do not overlap in energy (depending on the degree of electron doping). Such a superconducting state would likely be immune to nonmagnetic impurities~\cite{Chen_arXiv_2016}, as elastic scattering would not mix opposite-sign quasiparticles. Indeed, QPI measurements have confirmed that there is no electron-hole pocket scattering near the gap energy~\cite{Huang_PRL_2015, Fan_NatPhys_2015}. Thus, the defect experiments by Fan \textit{et al.} could also be consistent with ``incipient'' $s_{+-}$ gap symmetry. 

Recently, Huang \textit{et al.} uncovered a $\Gamma$ electron pocket 80 meV above $E_F$ (\textbf{Figure~\ref{FigBS}\textit{b}}), using empty-state STM measurements~\cite{Huang_PRL_2015}. This pocket may be similar to one discovered in K-coated bulk FeSe$_{0.55}$Te$_{0.45}$~\cite{Zhang_APL_2014} and FeSe~\cite{Wen_NatComm_2016}, or a shallow $Z$-electron Fermi pocket in 3D (Tl, Rb)$_y$Fe$_{2-x}$Se$_2$~\cite{Liu_PRL_2012}. Given that this pocket lies within the SrTiO$_3$ phonon energy (100 meV), it may be interesting to explore whether it has any positive contribution to $T_c$ in 1UC FeSe/SrTiO$_3$~\cite{Shi_arXiv_2016}. 

\section{SUMMARY AND OUTLOOK}

In this review, we have surveyed key experimental and theoretical developments related to 1UC FeSe/SrTiO$_3$ from its time of discovery, 2012, to early 2016. The major themes we have presented and developed can be captured in the following five statements: 

\begin{enumerate}
\item Monolayer FeSe on SrTiO$_3$ exemplifies a dramatic interface effect, in which a unit-cell layer of free-standing FeSe is non-superconducting down to 2.2 K, but subsequently exhibits $T_c$ ranging from 65 K to 109 K when coupled to SrTiO$_3$.
\item Experiments probing an air-sensitive, monolayer film are demanding in nature, but this challenge motivates the development and use of improved $\textit{in-situ}$ instrumentation, such as four-probe STM, which in turn may lead to discoveries of new systems.
\item Capping of 1UC FeSe/SrTiO$_3$ has not yet been optimized. Not only is it crucial for protecting films from atmospheric exposure, it simultaneously provides a second interface that could yet be engineered to enhance electronic properties.
\item Experiments have uncovered clues of both conventional and unconventional mechanisms of pairing in 1UC FeSe/SrTiO$_3$. While the ARPES replica bands are best explained by cross-interface coupling to SrTiO$_3$ phonon modes, more experiments verifying their nature and influence on $T_c$ are desirable~\cite{Zhang_arXiv_2016}. And although accumulated experience with iron-based superconductors might suggest the importance of spin (and orbital) fluctuations, direct measurements of magnetic excitations in 1UC FeSe/SrTiO$_3$ are still needed.
\item Various pairing scenarios have been proposed in which multiple bosons, such as phonons and spin fluctutations, can work cooperatively in a multiband environment to enhance $T_c$ in 1UC FeSe/SrTiO$_3$. Although these theories still require experimental confirmation, they represent an appealing ``best-of-both-worlds'' approach to finding and creating superconductors with even higher $T_c$. Combined with the layered 2D architecture of 1UC FeSe/SrTiO$_3$ in which these cooperative effects may be realized and engineered, many possibilies abound down the road. 
\end{enumerate}

\section*{DISCLOSURE STATEMENT}
The authors are not aware of any affiliations, memberships, funding, or financial holdings that might be perceived as affecting the objectivity of this review. 

\section*{ACKNOWLEDGMENTS}
The authors would like to thank C.-Z. Chang, S. Coh, S. Fang, J.-F. Ge, P. J. Hirschfeld, E. Kaxiras, S. A. Kivelson, D.-H. Lee, I. I. Mazin, Z.-X. Shen, C.-L. Song, T. A. Webb, and K. Zou for valuable discussions that have informed many of the views and ideas presented in this review. Our work on 1UC FeSe/SrTiO$_3$ was supported by the National Science Foundation under Grants No. DMR-0847433 and No. PHY-1231319 (STC Center for Integrated Quantum Materials), and the Gordon and Betty Moore Foundation's EPiQS Initiative through Grant. No. GBMF4536. J. E. Hoffman acknowledges support from the Canadian Institute for Advanced Research.


\end{document}